\documentclass[openacc]{rstransa}

\usepackage{amsmath}
\usepackage{amssymb}
\usepackage{natbib}
\usepackage{graphicx}
\usepackage{bm}
\usepackage[many]{tcolorbox}
\usepackage[shortlabels]{enumitem}

\newcommand*\aap{A\&A}

\newcommand*\apj{ApJ}
\newcommand*\apjl{ApJ}

\newcommand*\mnras{MNRAS}

\newcommand*\solphys{Sol.~Phys.}

\newcommand{\FIG}[1]{#1}

\titlehead{Research}

\begin{document}

\title{Radiative loss and ion-neutral collisional effects in astrophysical plasmas}

\author{
B. Popescu Braileanu$^{1}$, R. Keppens$^{1}$}

\address{$^{1}$ Centre for mathematical Plasma Astrophysics, Department of Mathematics, KU Leuven, Celestijnenlaan 200B, 3001 Leuven, Belgium }

\subject{Sun: filaments, prominences}


\corres{B. Popescu Braileanu\\
\email{beatriceannemone.popescubraileanu\\@kuleuven.be}}

\begin{abstract}
     In this paper we study the role of radiative cooling in a two-fluid model consisting of coupled neutrals and charged particles. 
We first analyze the linearized two-fluid equations where we include radiative losses in the energy equation for the charged particles.
In a 1D geometry for parallel propagation and in the limiting cases of weak and strong coupling, it can be shown analytically 
 that the instability conditions for the thermal mode and the sound waves, 
the isobaric and isentropic criteria,  respectively, remain unchanged with respect to  one-fluid radiative plasmas. 
For the parameters considered in this paper, representative for the solar corona,  the radiative cooling produces growth of the thermal mode and damping of the sound waves.
In the weak coupling limit, the growth of the thermal instability and the damping of the sound waves is as derived in \cite{field}
using the charged fluid properties. When neutrals are included and are sufficiently coupled to the charges, 
the thermal mode growth rate and the wave damping 
both reduce by the same factor, which depends on the ionization fraction only.
For a heating function which is constant in time, we find that the growth of the thermal mode and the damping
of the sound waves are slightly larger. 
The numerical calculation of the eigenvalues of the general system of equations in a 3D geometry confirm the 
analytic results. We then run 2D fully nonlinear simulations which give consistent results: a higher ionization fraction or lower coupling will increase the growth rate.
The magnetic field contribution is negligible in the linear phase. Ionization-recombination effects might play  an important role because 
the radiative cooling produces a large range of temperatures in the system. In the numerical simulation, 
after the first condensation phase, when the minimum temperature is reached,
the fraction of neutrals increases four orders of magnitude because of the recombination.  
\end{abstract}


\begin{fmtext}

\end{fmtext}


\maketitle

\section{Introduction}
\label{sec:intro}
Partial ionization effects are important in many astrophysical contexts. In the solar atmosphere ion-neutral collisions become important in the regions where a cool dense mostly neutral plasma changes to
a tenuous hot and ionized plasma. This is the case of the solar chromosphere, but also in prominence-corona transition regions. It is usually assumed that in prominences the ionization is collisional 
and the recombination is produced by radiative processes \citep{2019parenti}. 
At the same time, many recent works 
\citep[e.g.][]{coronalRainPatrick,niels,joris,veronika,xiaohong1,xiaohong2,jack,valeria,nicolas,yuhao,chun,antiochos,schr,klim,luna}
 support the original claim by \cite{parker} that thermal instability is a
fundamental process that can trigger in-situ condensations, like prominences and coronal rain in the solar atmosphere, but also denser clumps in galactic winds, or in other astrophysical contexts \citep{gronke,proga,waters}.
{ For this reason,  the linear mode which
grows exponentially due to the energy loss is usually called ``thermal'' or ``condensation'' mode \citep[see][]{field}.}
 Here, we will revisit the fundamental theory behind thermal instability initiated by \cite{parker} and \cite{field}, but this time extend the findings to neutral-plasma mixtures, which play a role for especially the finer-scale details of the prominence or coronal rain dynamics \citep{prompi,rti1,rti2,rti3}.

It was shown that the magnetic field does not affect significantly the behaviour of the thermal mode \citep{field}.
From the linear analysis the motion in the condensation mode is mostly along the field lines, therefore they do not produce perturbations of the magnetic field. 
It was shown analytically that, when the propagation is not exactly perpendicular, the magnetic field has no effect on the linear growth of the thermal mode and the magnetic field with magnitude at equipartition value is enough to 
align the small motions of the condensation mode with the field \citep{field}. 
However, for { exactly} perpendicular propagation, it is shown analytically that the perturbation in the magnetic pressure 
is not zero and has a slight stabilizing effect on the linear growth of the thermal mode \citep{field}.
 The non-linear evolutions show systematic field-guided motions of the condensations \citep{niels}. The shape and evolution  of the condensations
 depend on the radiative cooling curve used \citep{joris}.

The effect of partial ionization on wave propagation in a uniform atmosphere has been studied by several authors \citep{2013Soler,2013SolerB}.  The authors find cutoff values for the Alfv\'en and magnetoacoustic waves because of the frictional force competing with the restoring force.
In these works the dispersion relation obtained { analytically was studied numerically} and some simplifications were done, such as including the ion-neutral coupling only in the momentum equation. We here augment these findings with an emphasis on thermal instabilities, and by allowing both thermal (energy) and momentum coupling.
{ The partial ionziation effects were studied in the dynamics of falling blobs, with properties characteristic for coronal rain \citep{coronalRain1,coronalRain2}, however,
 the  radiative cooling effect was not included and the formation of these blobs was not considered in their study.}

In ion-neutral mixtures, one should consider ionization-recombination aspects as well. They may enter especially during non-linear evolutions, as it was found that
ionization-recombination processes do not modify the linear growth of the Rayleigh-Taylor instability, but during the non-linear phase the neutral drops become surrounded by a layer of ionized material \citep{rti2}.
It was also shown that ionization and recombination processes alter significantly the structure of a two-fluid slow shock \citep{ionrecBen}. Here, we will consider ionization-recombination effects in non-linear evolutions of thermally unstable setups.
The very large temperatures in the solar corona makes the gravitational scale height very large, of several hundreds of megameters for a temperature of one million K. Therefore, for the study of scales similar to one megameter, neglecting gravity and considering the density and pressure uniform is a good approximation. This simplifies the setups we use for our nonlinear studies.

In this work we include the thermal exchange { between charges and neutrals in both energy equations}, as well as radiative cooling in the charged energy equation. We  solve the dispersion
relation numerically and we study the effect of the collisions and the radiative losses on 
damping the waves and the growth of the thermal mode. We then test the linear analytical results with simulations where we can also study the non-linear evolution. 
In order to see the development of the two-fluid thermal instability in the nonlinear phase, 
we perform fully nonlinear 2D, two-fluid simulations, using the same uniform background.

\section{Governing equations}
\label{seq:eqs}

The radiative cooling term is in general a function of temperature and density, $F(\rho,T)$. If we consider
a background atmosphere with density $\rho_0$ and temperature $T_0$ in thermal equilibrium, 
the radiative losses have to be compensated by an unspecified heating function so that: 
\begin{equation}
F(\rho_0, T_0) = 0\,.
\end{equation}
Usually, this term is written as \citep[see ][]{field}, 
\begin{equation}
F(\rho, T) = \rho \mathcal{L}(\rho,T)\,,
\end{equation}
where $\mathcal{L}(\rho,T)$ is a generalized heat-loss function which fulfills:
\begin{equation}
\mathcal{L}(\rho_0,T_0) = 0 
\end{equation}
for the thermal equilibrium condition, and it is usually considered to be 
for the optically thin radiative loss prescription:
\begin{equation}
\mathcal{L}(\rho_0,T_0) = \rho \Lambda(T) - \rho_0 \Lambda_0\,,
\end{equation}
where the cooling curve, $\Lambda$, is a function of temperature only. The subindex $_0$ means that the quantity has to be evaluated for
the background atmosphere variables.
Therefore
the contribution of the radiative cooling terms in the energy equation becomes:
\begin{equation}
\label{RCf}
F(\rho, T) = \rho \left[\rho  \Lambda - \rho_0 \Lambda_0 \right]\,,
\end{equation}
thus having a time-varying heating function which  depends on the time-varying density, 
rather than on equilibrium variables only.
In our simulations, however, we considered a constant heating function \citep[see ][]{joris,niels}, having the contribution of the radiative losses,
when the background is in thermal equilibrium actually given by
\begin{equation}
\label{RCc}
F(\rho, T)  = \rho^2 \Lambda - \rho_0^2 \Lambda_0\,.
\end{equation}

\noindent
The nonlinear two-fluid equations evolved by the code {\tt MPI-AMRVAC} \citep{amrvac}, when we split the variables \citep[see ][]{nitin} into time-dependent { (subscript ``$_1$'')}
and time-independent variables { (subscript ``$_0$'')} for neutral and charged densities, neutral and charged pressures, and the magnetic field, become with the splitting obeying
\begin{eqnarray}
\rho_{\rm n} =  \rho_{\rm n0} + \rho_{\rm n1}\,,
\rho_{\rm c} =  \rho_{\rm c0} + \rho_{\rm c1}\,,
p_{\rm n} =  p_{\rm n0} + p_{\rm n1}\,,
p_{\rm c} =  p_{\rm c0} + p_{\rm c1}\,,
\mathbf{B} =  \mathbf{B}_{\rm 0} + \mathbf{B}_{\rm 1}\,,\nonumber\\
\frac{\partial \rho_{\rm n0}}{\partial t}=0\,,
\frac{\partial \rho_{\rm c0}}{\partial t}=0\,,
\frac{\partial p_{\rm n0}}{\partial t}=0\,,
\frac{\partial p_{\rm c0}}{\partial t}=0\,,
\frac{\partial \mathbf{B}_{0}}{\partial t}=0\,,
\end{eqnarray}
as follows:
\begin{eqnarray}
\label{eqs:2fl_start}
\frac{\partial \rho_{\rm n1}}{\partial t} + \nabla \cdot \left(\rho_{\rm n}\mathbf{v}_{\rm n}\right) = 
S_{\rm n}\,,
 \\
\frac{\partial \rho_{\rm c1}}{\partial t} + \nabla \cdot \left(\rho_{\rm c}\mathbf{v}_{\rm c}\right) = 
-S_{\rm n} \,,
\\
\frac{\partial (\rho_{\rm n}\mathbf{v_{\rm n}})}{\partial t} + \nabla \cdot \left(\rho_{\rm n}\mathbf{v_{\rm n}} \mathbf{v_{\rm n}}  + p_{\rm n1} \mathbb{I}  \right)
 = \rho_{\rm n1}\mathbf{g} +
\mathbf{R}_{\rm n} \,,
\\
\frac{\partial (\rho_{\rm c}\mathbf{v_{\rm c}})}{\partial t} + \nabla \cdot \left[\rho_{\rm c}\mathbf{v}_{\rm c} \mathbf{v}_{\rm c} +  
\left( p_{\rm c1} +  \frac{B_1^2}{2} + \mathbf{B_0}\cdot \mathbf{B_1}\right) \mathbb{I}- \mathbf{B_1} \mathbf{B_1}- \mathbf{B_1} \mathbf{B_0}- \mathbf{B_0} \mathbf{B_1}
 \right]\nonumber\\[0.12cm]
 = \rho_{\rm c1}\mathbf{g}  
-\mathbf{R}_{\rm n} \,,
\\
\frac{\partial}{\partial t} \left(e_{\rm n1}+\frac{1}{2}\rho_{\rm n} \mathbf{v}_{\rm n}^2\right)   
+  \nabla \cdot \left[ \mathbf{v}_{\rm n} \left(e_{\rm n}+\frac{1}{2}\rho_{\rm n} \mathbf{v}_{\rm n}^2 + p_{\rm n1}\right)    \right ]  
 = \rho_{\rm n1}\mathbf{v}_{\rm n}\cdot \mathbf{g} -p_{\rm n0} \nabla \cdot \mathbf{v}_{\rm n} + 
M_{\rm n} \,,
\\[0.15cm]
\frac{\partial}{\partial t} \left(  e_{\rm c1}+\frac{1}{2}\rho_{\rm c} \mathbf{v}_{\rm c}^2  + \frac{B_1^2}{2}\right) +  \nabla \cdot \left[ \mathbf{v}_{\rm c} \left(e_{\rm c} + 
\frac{1}{2}\rho_{\rm c} \mathbf{v}_{\rm c}^2+ p_{\rm c1} \right)  -  (\mathbf{v}_{\rm c} \times \mathbf{B})\times\mathbf{B_1}  \right ]   &  
 \nonumber\\[0.12cm]
= \rho_{\rm c1}\mathbf{v}_{\rm c}\cdot \mathbf{g} 
-p_{\rm c0} \nabla \cdot \mathbf{v}_{\rm c} + \mathbf{J_0}\cdot (\mathbf{v}_{\rm c} \times \mathbf{B}_1) 
-\rho_c^2 \Lambda + \rho_{\rm c0}^2 \Lambda_0
-M_{\rm n} \,,
\label{eqs:encharge}
\\
\label{eqs:2fl_end}
\frac{\partial \mathbf{B_1}}{\partial t} = \nabla \times  (\mathbf{v}_{\rm c} \times \mathbf{B})\,. 
\end{eqnarray}
{ The above equations are written in non-dimensional units. $\mathbb{I}$ is the identity matrix.}
The equations contain velocity fields for neutrals and charges, $\mathbf{v}_{n}$ and $\mathbf{v}_{c}$. Terms
$S_n$, $\mathbf{R_n}$ and $M_n$ in the above equations are the collisional coupling terms, implemented as described by 
Eqs.~(\ref{eq:collts})-(\ref{eq:collte}), shown in Appendix~\ref{sec:collt}, being the same
Eqs.~(13) in \cite{paper2f}, which introduce a basic coupling parameter called $\alpha$, which quantifies collisional coupling. 
We considered the radiative losses in the charged energy equation Eq.~(\ref{eqs:encharge}).
In the two-fluid model we use the same radiative losses curves which define the profile of $\Lambda$ as a function of temperature,
implemented for the MHD model \citep{joris}, now calculated using the charged fluid temperature, $\Lambda=\Lambda(T_{\rm c})$.
In the above equations it is assumed that the time-independent background is in mechanical equilibrium, 
fulfilling the magneto-hydrostatic equilibrium for the charged fluid and the 
hydrostatic equilibrium for the neutral fluid, meaning uniform charges and neutral pressures  and force-free field (i.e. $\mathbf{J}_0\times\mathbf{B}_0$) when there is no gravity (gravity is incorporated in the equations as implemented above where gravitational acceleration is $\mathbf{g}$, but we ignore it further on). Our background magnetic field will even be a simple uniform field $\mathbf{B}_0$ in what follows.
We also assume that the background is in thermal equilibrium. 
We have the same background temperature for neutrals and charges, $T_{\rm n0}=T_{\rm c0}=T_0$, and the radiative losses
of the equilibrium charged fluid is compensated by a heating mechanism, constant in time, which is explicitly removed from  
Eqs.~(\ref{eqs:2fl_start})-(\ref{eqs:2fl_end}) solved by code. Note that we did not incorporate anisotropic thermal conduction, and we adopt a simple ideal-gas-law closure for pressures $p$, internal energies $e$ and temperatures $T$.

\section{Linear approach}
\label{sec:lin}
The time evolution of a set of small perturbations $\mathbf{u}$ of a stationary  background 
is assumed to be $\propto \text{exp}(i\omega t)$ and the equations are linearized, becoming
\begin{equation}
\label{eq:eig}
\left( \hat{\mathbf{M}}-\omega\mathbb{I}\right) \mathbf{u}=0\,,
\end{equation}
where the elements of the complex matrix $\hat{\mathbf{M}}$ depend on the background values.
The background has  densities  $\rho_{\rm c0}$, $\rho_{\rm n0}$,
and pressures $p_{\rm c0}$ and $p_{\rm n0}$, of charges and neutrals, respectively and a uniform magnetic field with magnitude $B_0$. 
{
We introduce some quantities, the total density and presure, the sound speed for neutrals, charges and total fluid,
 as well as the Alfv\'en, fast and slow speeds of the charged fluid and the total fluid,   calculated for the background values:
\begin{eqnarray}
\label{eq:rhot0}
\rho_0=\rho_{\rm c0} + \rho_{\rm n0}\,,\quad p_0=p_{\rm c0} + p_{\rm n0} \\
\label{eq:cn0}
c_{\rm n0} = \sqrt{\gamma\frac{p_{\rm n0}}{\rho_{\rm n0}}}\,,\\
\label{eq:cc0}
c_{\rm c0} = \sqrt{\gamma\frac{p_{\rm c0}}{\rho_{\rm c0}}}\,,\\
\label{eq:ct0}
c_{\rm t0} = \sqrt{\gamma\frac{p_0 }{\rho_0}}\,,\\
\label{eq:vac0}
v_{\rm Ac0} = \frac{B_0}{\sqrt{\rho_{\rm c0}}}\,,\\
\label{eq:vat0}
v_{\rm At0} = \frac{B_0}{\sqrt{\rho_0}}\,,\\
\label{eq:fc0}
f_{\rm c0}=\sqrt{\frac{1}{2}\left[ v_{\rm Ac0}^2 + c_{\rm c0}^2 + \sqrt{ \left(v_{\rm Ac0}^2 + c_{\rm c0}^2 \right)^2 - 4  v_{\rm Ac0}^2  c_{\rm c0}^2  \text{cos}^2(\theta)}  \right]}\,,\\
\label{eq:ft0}                                                                                                                                                      
f_{\rm t0}=\sqrt{\frac{1}{2}\left[ v_{\rm At0}^2 + c_{\rm t0}^2 + \sqrt{ \left(v_{\rm At0}^2 + c_{\rm t0}^2 \right)^2 - 4  v_{\rm At0}^2  c_{\rm t0}^2  \text{cos}^2(\theta)}  \right]}\,,\\
\label{eq:sc0}                                                                                                                                                      
s_{\rm c0}=\sqrt{\frac{1}{2}\left[ v_{\rm Ac0}^2 + c_{\rm c0}^2 - \sqrt{ \left(v_{\rm Ac0}^2 + c_{\rm c0}^2 \right)^2 - 4  v_{\rm Ac0}^2  c_{\rm c0}^2  \text{cos}^2(\theta)}  \right]}\,,\\
\label{eq:st0}                                                                                                                                                      
s_{\rm t0}=\sqrt{\frac{1}{2}\left[ v_{\rm At0}^2 + c_{\rm t0}^2 - \sqrt{ \left(v_{\rm At0}^2 + c_{\rm t0}^2 \right)^2 - 4  v_{\rm At0}^2  c_{\rm t0}^2  \text{cos}^2(\theta)}  \right]}\,.
\end{eqnarray}
}

\noindent
{
In a 3D cartesian geometry, without  loss of generality, we choose the magnetic field $\mathbf{B_0}$ in the $xz$ plane and the direction of propagation in the $x$-direction.
For a uniform background, the perturbations from Eq.~(\ref{eq:eig}) can be written as:
}
\begin{equation}
\label{eq:pert}
\mathbf{u}= (\rho_{\rm c1},\rho_{\rm n1}, v_{\rm cx},v_{\rm nx}, v_{\rm cy},v_{\rm ny},v_{\rm cz}, 
v_{\rm nz}, {p_{\rm c1}}, {p_{\rm n1}}, {B_{\rm x1}}, {B_{\rm y1}}, {B_{\rm z1}}  )^T
\text{exp}\left[i(\omega t - k_x x)\right]\,,
\end{equation}
{
$k_x$ is the wavenumber or spatial frequency and $\omega$ represents the temporal frequency. In our study,  $k_x$ is a real quantity
and $\omega$ is complex. 
}
The linearized two-fluid equations  for a  static background become:
\begin{eqnarray}
\label{eqs:lin3d}
\omega \rho_{\rm c1} - k_x \rho_{\rm c0} v_{\rm cx}= 0 \,,\nonumber\\
\omega \rho_{\rm n1} - k_x \rho_{\rm n0} v_{\rm nx}= 0 \,,\nonumber\\
i \omega \rho_{\rm c0} v_{\rm cx} -i k_x  p_{\rm c1} - 
i k_x B_{\rm z0}  B_{\rm z1}  + \alpha \rho_{\rm c0} \rho_{\rm n0} \left(v_{\rm cx} - v_{\rm nx}\right) =0\,,\nonumber\\\
i \omega \rho_{\rm n0} v_{\rm nx} -i k_x  p_{\rm n1}  -\alpha \rho_{\rm c0} \rho_{\rm n0} \left(v_{\rm cx} - v_{\rm nx}\right)=0 \,,\nonumber\\\
i \omega \rho_{\rm c0} v_{\rm cy}  + i k_x B_{\rm y1}  B_{\rm x0}+\alpha \rho_{\rm c0} \rho_{\rm n0} \left(v_{\rm ny} - v_{\rm cy}\right)=0 \,,\nonumber\\\
i \omega \rho_{\rm n0} v_{\rm ny} -\alpha \rho_{\rm c0} \rho_{\rm n0} \left(v_{\rm ny} - v_{\rm cy}\right)=0 \,,\nonumber\\\
i \omega \rho_{\rm c0} v_{\rm cz}  + i k_x B_{\rm z1}  B_{\rm x0}  + \rho_{\rm c0} \alpha \rho_{\rm n0} (v_{\rm cz} - v_{\rm nz})=0
\,,\nonumber\\\
i \omega \rho_{\rm n0} v_{\rm nz}  - \alpha \rho_{\rm c0} \rho_{\rm n0} (v_{\rm cz} - v_{\rm zn})=0
\,,\nonumber\\\
i \omega p_{\rm c1}  - i c_{\rm c0}^2 \omega \rho_{\rm c1} +  
(\gamma-1) \left[ \frac{1}{2} \left(\rho_{\rm c0} p_{\rm c1} - p_{\rm c0} \rho_{\rm c1}\right) \frac{d\Lambda_0}{dT} + 
2 \rho_{\rm c0} \rho_{\rm c1} \Lambda_0\right] \nonumber\\ 
+
 \alpha \left[-\rho_{\rm c1} p_{\rm n0} - \rho_{\rm c0} p_{\rm n1} + \frac{1}{2} (\rho_{\rm n1} p_{\rm c0} +  \rho_{\rm n0} p_{\rm c1})\right]=0
\,,\nonumber\\\
i \omega p_{\rm n1}  - i c_{\rm n0}^2  \omega \rho_{\rm n1}   - 
\alpha \left[-\rho_{\rm c1} p_{\rm n0} - \rho_{\rm c0} p_{\rm n1} + \frac{1}{2} (\rho_{\rm n1} p_{\rm c0} +  \rho_{\rm n0} p_{\rm c1})\right] =0
\,,\nonumber\\\
B_{\rm x1} = 0\,,\nonumber\\
B_{\rm y1} \omega + B_{\rm x0} k_x v_{\rm cy}=0 \,,\nonumber\\
B_{\rm z1} \omega  - B_{\rm z0} k_x v_{\rm cx} + B_{\rm x0} k_x v_{\rm cz} = 0\,.
\end{eqnarray}
where the unknowns are the complex amplitudes of the perturbations, as shown in Eq.~(\ref{eq:pert}).
$\gamma$ is the ratio of specific heats.

We believe that, similarly to the case of the Rayleigh-Taylor instability \citep{rti2}, the ionization/recombination processes are not important in the linear phase, and ignored them here to keep the model simpler.
While we neglected ionization/recombinations in the two-fluid model in the linear analysis,  
we included the elastic collisions in the momentum equations and the thermal exchange in the energy
equations, i.e. all terms containing the coupling parameter $\alpha$.  

The linearized form of $F(\rho, T)$ from Eq.~(\ref{RCc})
around background density and temperature values $\rho_0$ and $T_0$ is:
\begin{equation}
F(\rho, T) = \frac{\partial F_0}{\partial \rho} \rho_1 +  \frac{\partial F_0}{\partial T} T_1\,.
\end{equation}
In the linear assumption, for a generic equation of state written in non-dimensional form as $p = A \rho T$,  where $A$ is the 
inverse of the non-dimensional mean molecular weight \citep[see also Eq.~(14) in][]{field},
\begin{equation}
T_1 = \frac{1}{A \rho_0} \left( p_1 - p_0 \frac{\rho_1}{\rho_0} \right)\,.
\end{equation} 
This gives the contribution of the radiative cooling term in the linearized equation for $F$ defined in Eq.~(\ref{RCc}) :
\begin{equation}
\label{eq:temp1}
F(\rho, T)  = 2 \rho_0 \Lambda_0 \rho_1 +  \frac{1}{A} \frac{\partial \Lambda_0}{\partial T} (p_1 \rho_0 - p_0 \rho_1)\,.
\end{equation}
The difference from considering  a constant heating function instead of a time-varying heating function, 
having  $F$ as defined in Eq.~(\ref{RCc}) rather than the definition in Eq.~(\ref{RCf}),  is the 
factor of 2 multiplying the term containing $\Lambda_0$, 
compared to Eq.~(13) in \cite{field} and Eq.~(9) in \cite{niels}.
We will see how the analytic expressions are slightly modified by this choice, as we will highlight this factor 2 in the following derivation.
We used the results for the constant heating function in our analysis in order to have full consistency with the equations solved by the code.

In the two-fluid model, the radiative cooling (RC) term is added in the charges energy equation Eq.~(\ref{eqs:encharge}),
similarly to MHD, but using charged fluid density and temperature in the evaluation of the term. 
The RC is considered for high temperature plasmas optically thin, where the plasma is fully ionized, so that
the radiative losses are related to the charged particles. 
We assume Hydrogen only and charge neutrality, therefore $A=2$ for the charged fluid in the above Eq.~(\ref{eq:temp1}).

The collisional parameter $\alpha$ depends weakly on plasma parameters, being proportional to the square root of the average temperature between neutrals and charges. It is assumed constant, calculated using equilibrium values,  in the above linearized  Eqs.~(\ref{eqs:lin3d}). 
 
The system Eqs.~(\ref{eqs:lin3d})  has 13 equations, but the $x$-induction equation  does not contain $\omega$ 
because of the additional constraint of zero magnetic field divergence (in the linear assumption one of the induction equations 
is equivalent to the zero divergence of the magnetic field),  therefore there  are 12 solutions for the eigenvalue problem Eq.~(\ref{eq:eig}). 
When there is no coupling between the charges and neutrals, there are
five neutral modes and seven MHD modes divided in
two trivial entropy modes, six propagating modes for the charged fluid: fast, slow and Alfv\'en; two propagating sound modes and two trivial shear modes for the neutral fluid \citep{bookmhd}.

We can also observe in the Eqs.~(\ref{eqs:lin3d}) that the $y$-momentum equations for neutrals and charges and the $y$-induction equation are decoupled from the rest.
Therefore the  Alfv\'en branch decouples:
\begin{equation}
\label{eq:alfbr}
B_{\rm x0}^2 k_x^2 (\omega - i \alpha \rho_{\rm c0}) - \omega^2 \rho_{\rm c0} \left[\omega - i \alpha (\rho_{\rm c0}+\rho_{\rm n0})\right]=0\,,
\end{equation}
{ and } the non-ideal terms in the energy equations (the thermal exchange and radiative cooling) do not affect the Alfv\'en waves.
Eq.~(\ref{eq:alfbr}) is equivalent to Eq.~(14) from \cite{2013Soler} and Eq.~(39) in \cite{beatrice1}.
The remaining equation, of 9th order is not shown here  because of its complicated form, however with some additional assumptions it particularizes into some known forms.
When the radiative cooling is not taken into account, the charges entropy mode  decouples, giving the root $\omega=0$.

The angle between the wavevector $\mathbf{k}$ and $\mathbf{B_0}$  can equivalently be described using the following transformations:
\begin{equation}
k_x=k\,,
B_{\rm z0}=\frac{k_\perp}{k} B_0\,,
B_{\rm x0}=\frac{k_\parallel}{k} B_0\,,
k_{\perp}=k \text{sin}(\theta)\,,
k_{\parallel}=k \text{cos}(\theta)\,,
\end{equation}
where we introduce field-aligned parallel and field-perpendicular wave numbers. 
We will now discuss various limits of the dispersion relation.

\subsection{The case of purely perpendicular propagation and no radiative cooling}

In this case, { regardless the value of $k$,} we obtain a triple trivial root $\omega=0$, a purely damped root $\omega=i \alpha (\rho_{\rm c0} + \rho_{\rm n0})$ and the remaining fifth order dispersion relation becomes
\begin{eqnarray}
    v_{\rm Ac0}^2 k^2\left[-2 \omega^3 - i \alpha k^2 p_{\rm c0} + \alpha^2 \omega \rho_{\rm c0} (2 \rho_{\rm c0} + \rho_{\rm n0}) + i \alpha \omega^2 (4 \rho_{\rm c0} + \rho_{\rm n0}) + c_{\rm n0}^2 k^2 (2\omega - i \alpha \rho_{\rm n0})\right] + \nonumber\\
    c_{\rm c0}^2 k^2 \left[2 c_{\rm n0}^2 k^2 \omega - 2 \omega^3 - i \alpha k^2 p_{\rm c0} + 4i \alpha \omega^2 \rho_{\rm c0} + \alpha^2 \omega \rho_{\rm c0} (2 \rho_{\rm c0} + \rho_{\rm n0})\right] + \nonumber \\ 
    c_{\rm n0}^2 k^2 \left[-2 \omega^3 - 2 i \alpha k^2 p_{\rm n0} + 3 i \alpha \omega^2 \rho_{\rm n0} + \alpha^2 \omega \rho_{\rm n0} (2 \rho_{\rm c0} + \rho_{\rm n0})\right] + \nonumber \\
    \omega^2 (2 \omega^3 + i\alpha k^2 (p_{\rm c0} + 2 p_{\rm n0}) - i \alpha \omega^2 (4 \rho_{\rm c0} + 3 \rho_{\rm n0}) - \alpha^2 \omega (2 \rho_{\rm c0}^2 + 3 \rho_{\rm c0} \rho_{\rm n0} + \rho_{\rm n0}^2))=0 \,.
\end{eqnarray} 
We { used the equilibrium charged fluid Alfv\'en speed, the neutral and charged fluid sound speeds, 
introduced in Eqs.~(\ref{eq:vac0}), (\ref{eq:cn0}), (\ref{eq:cc0}), respectively,}
 and $k\equiv k_\perp$.
When the coupling is taken into account only in the momentum equations, 
but not in the energy equation (i.e. when the thermal exchange is neglected),
another root $\omega=0$ is obtained and the then remaining fourth order dispersion relation reduces to
\begin{eqnarray}
 (c_{\rm c0}^2 + v_{\rm Ac0}^2) c_{\rm n0}^2 k^4 + \omega^4 - (c_{\rm c0}^2 + v_{\rm Ac0}^2) k^2 \omega (\omega - i \alpha \rho_{\rm c0}) - i \alpha \omega^3 (\rho_{\rm c0} + \rho_{\rm n0}) - c_{\rm n0}^2 k^2 \omega (\omega - i \alpha \rho_{\rm n0})=0\,.
\label{eq:dispFast}
\end{eqnarray} 
This represents two pairs of neutral sound and charges fast waves, as expected, modified by momentum coupling terms proportional to $\alpha$.
The above Eq.~\ref{eq:dispFast} is exactly Eq.~(56) from \cite{beatrice1}, where the terms related to stratification $b_n=b_c=0$.

\subsection{The case of purely parallel propagation}
Assuming parallel propagation $k\equiv k_\parallel$ to the magnetic field ($\theta=0$), the 9th order dispersion relation can be factorized
into a term which represents the Alfv\'en branch: 
\begin{equation}
B_0^2k^2(\omega - i \alpha\rho_{\rm c0}) - \omega^2\rho_{\rm c0} \left[\omega - i \alpha(\rho_{\rm c0} + \rho_{\rm n0})\right]=0\,.
\label{eq:sound2perp}
\end{equation}
and a 6th order dispersion relation which contains the 1D sound branch. 
As  the magnetic field does not affect significantly the behaviour of the thermal mode \citep{field}, we will see that this simplification 
is a very good approximation. 
We will consider further some limiting cases and compare the results to others existing in the literature.

\begin{itemize}
\item {\bf No RC}

When RC is not taken into account,
the root $\omega=0$, corresponding to the charges entropy mode,  appears in the 6th order dispersion relation, and the remaining 5th order dispersion relation of the sound branch for 
parallel propagation becomes: 
\begin{eqnarray}
c_{\rm c0}^2k^2\left[2 c_{\rm n0}^2k^2\omega - 2\omega^3 - i \alpha k^2p_{\rm c0} + 4i \alpha\omega^2\rho_{\rm c0} + \alpha^2\omega\rho_{\rm c0}(2\rho_{\rm c0} + \rho_{\rm n0})\right] + \nonumber\\ 
c_{\rm n0}^2k^2\left[-2\omega^3 - 2i \alpha k^2 p_{\rm n0} + 3i \alpha\omega^2\rho_{\rm n0} + \alpha^2\omega\rho_{\rm n0}(2\rho_{\rm c0} + \rho_{\rm n0})\right] + \nonumber \\ 
\omega^2 \left[2\omega^3 + i \alpha k^2(p_{\rm c0} + 2 p_{\rm n0}) - i \alpha\omega^2(4\rho_{\rm c0} + 3\rho_{\rm n0}) - \alpha^2\omega(2\rho_{\rm c0}^2 + 3\rho_{\rm c0}\rho_{\rm n0} + \rho_{\rm n0}^2)\right] = 0\,.
\end{eqnarray}
When the ion-neutral coupling is taken into account only in the momentum equation, there is another root $\omega=0$ and the remaining 4th order dispersion relation
\begin{eqnarray}
c_{\rm c0}^2c_{\rm n0}^2k^4 + \omega^4 - c_{\rm c0}^2k^2\omega (\omega  -  i \alpha \rho_{\rm c0}) -  i \alpha \omega^3(\rho_{\rm c0} + \rho_{\rm n0}) - c_{\rm n0}^2k^2\omega (\omega  -  i \alpha \rho_{\rm n0})=0\,.
\label{eq:sound2par}
\end{eqnarray}
represents the two pairs (neutrals and charges) of sound waves.
We can observe how Eq.~(\ref{eq:sound2par})  is similar to the above Eq.~(\ref{eq:sound2perp}), the only difference being that the charges sound speed was replaced by the fast magneto-acoustic speed.
The above equation is identical to Eq.~(31) from \cite{beatrice1}.

\item {\bf Limits depending on coupling regime}

In the weak coupling regime ($\alpha\to0$), the two-fluid 1D sound branch
 dispersion relation is still 6th order, but now there is a trivial root $\omega=0$, and the neutral sound branch, unaffected by the RC decouples into
\begin{equation}
\omega^2-k^2 c_{n0}^2=0 \,,
\end{equation}
and the remaining equation of 3rd order, 
when we replace $\omega=-i n$, so that we get 
a third order polynomial equation with real coefficients, which has exact analytic solutions, becomes:
\begin{equation}
\label{eq:spagen}
n^3 + a_2 n^2 + a_1  n + a_0 =0\,.
\end{equation}
{ Using the fact that the equilibrium temperature for neutrals and charges is the same, equal to $T_0$}, the coefficients $a_0$, $a_1$ and $a_2$ can be written as: 
\begin{eqnarray}
\label{eq:spa0}
a_2 = (\gamma-1) \rho_{\rm c0}^2 \frac{\partial \Lambda_0}{\partial T} T_0 \frac{1}{p_{\rm c0}}\,, \nonumber \\
a_1 = k^2 c_{\rm c0}^2\,,\nonumber\\
a_0 = (\gamma-1)k^2 \rho_{\rm c0}  \left[\frac{\partial \Lambda_0}{\partial T} T_0 - 2\Lambda_0 \right]\,,
\end{eqnarray}
being identical to Eq.~(15) in \cite{field}, if we neglect the thermal conductivity effects and we further replace 
$\frac{\partial \Lambda_0}{\partial T}=\frac{\mathcal{L}_T}{\rho_{\rm c0}}$
and $\Lambda_0=\frac{1}{2}\mathcal{L}_\rho$ (the factor 2 comes from considering the constant heating function). 

For strong coupling regime ($\alpha\to\infty$), the higher order terms in the 6th order dispersion relation
for parallel propagation disappear, and after removing a root $\omega=0$, a third order dispersion relation remains, 
Eq.~(\ref{eq:spagen}), but with different coefficients:
\begin{eqnarray}
\label{eq:spainf}
a_2 = (\gamma-1) \rho_{\rm c0}^2 \frac{\partial \Lambda_0}{\partial T} T_0 \frac{1}{p_{\rm 0}}\,,\nonumber  \\
a_1 = k^2 c_{\rm 0}^2\,,\nonumber\\
a_0 = (\gamma-1)k^2 \rho_{\rm c0}  \left[\frac{\partial \Lambda_0}{\partial T} T_0 - 2\Lambda_0 \right] \frac{ \rho_{\rm c0}}{\rho_0}\,,
\end{eqnarray}
where ${p_{\rm 0}}$, ${\rho_0}$, and $c_{\rm 0}$ are the pressure, density, and sound speed of the total fluid, 
therefore being consistent with those shown in Eqs.~(\ref{eq:spa0}).
In order to analyze the properties of the solutions for both cases, 
similarly to the derivation in \cite{field}, we can further apply the following transformations:
\begin{eqnarray}
\label{eq:sigma0}
y = \frac{n}{k c_{\rm c0}}\,,\nonumber \\
\sigma_T = \frac{(\gamma-1) \rho_{\rm c0}}{2 k c_{\rm c0}} \frac{\partial \Lambda_0}{\partial T}   \,,\nonumber \\
\sigma_\rho = \frac{(\gamma-1) \rho_{\rm c0} }{2 k c_{\rm c0} T_0}  (2\Lambda_0)\,,
\end{eqnarray}
for $\alpha \to 0$, and
\begin{eqnarray}
\label{eq:sigmaInf}
y = \frac{n}{k c_{\rm 0}}\,\nonumber\\
\sigma_T = \frac{(\gamma-1) \rho_{\rm c0}}{2 k c_{\rm 0}} \frac{\partial \Lambda_0}{\partial T} \frac{p_{\rm c0}}{p_{\rm 0}}  \,,\nonumber\\
\sigma_\rho = \frac{(\gamma-1) \rho_{\rm c0} }{2 k c_{\rm 0} T_0}  (2\Lambda_0) \frac{p_{\rm c0}}{p_{\rm 0}}\,,
\end{eqnarray}
for $\alpha \to \infty$.
The factor 2 in front of $\Lambda_0$ in the definition of $\sigma_\rho$ comes from having a constant heating function, { i.e.
having the radiative cooling contribution as from Eq.~(\ref{RCc}), rather than Eq.~(\ref{RCf})}.

$\sigma_T$ and $\sigma_\rho$ are slightly different for the $\alpha \to \infty$ compared to the $\alpha \to 0$ case,
being multiplied by a factor of $\frac{p_{\rm c0}}{p_{\rm 0}}$ and 
using the definition of the sound speed of the total fluid.
With these transformations we get an equation identical to Eq.~(18) in \cite{field} for both the case $\alpha \to 0$ and $\alpha \to \infty$, 
so that we can apply directly the same stability analysis.

Since the coefficients $\sigma_T$ and $\sigma_\rho$ defined in Eq.~(\ref{eq:sigmaInf}) for the strong coupling regime  
are proportional to those defined
in Eq.~(\ref{eq:sigma0}) for the weak coupling regime, the conditions to have 
one real positive root and two complex conjugate roots with negative real part
(meaning growth for the thermal mode and damping for the two propagating  sound waves)  are identical for both regimes
and are the isobaric criterium for instability  and the isentropic criterium for stability, defined by Eqs.~(23) and (24) in \cite{field}:
\begin{eqnarray}
\label{eq:crit}
\sigma_T - \sigma_\rho < 0   \,,\nonumber\\
\sigma_T + \frac{1}{\gamma-1} \sigma_\rho > 0  \,,
\end{eqnarray}
which become, after replacing $\sigma_T$ and $\sigma_\rho$:
\begin{eqnarray}
T_0 \frac{\partial \Lambda_0}{\partial T} - 2 \Lambda_0 <0 \,,\nonumber\\
T_0 \frac{\partial \Lambda_0}{\partial T} + \frac{2}{\gamma-1} \Lambda_0 > 0\,,
\end{eqnarray}
for both coupling regimes. The only difference is the factor 2 in front of $\Lambda_0$
which comes from having the constant heating function as already mentioned. 

The growth of the thermal mode and the damping of the sound waves in the high $k$ (short wavelength) regime can be approximated as 
the real part of the  leading term of Eq.~(30) in \cite{field}, 
\begin{eqnarray}
\label{eq:angrowth_n}
\tilde{g} =  \frac{\sigma_\rho -  \sigma_T}{ \gamma}\,,\\
\label{eq:andamp_n}
\tilde{\delta} = \frac{\sigma_\rho + (\gamma-1)  \sigma_T}{2  \gamma}\,.
\end{eqnarray}

The growth rate of the thermal mode goes to zero when $k\to 0$, except for the case  when $\sigma_T<0$ \citep{field}.
In that latter case, the growth can be  approximated by the leading term in Eq.~(33) from \cite{field}:
\begin{equation}
\label{eq:angrowth_smallk_n}
 \tilde{g}^L=-\sigma_T\,.
\end{equation}

These are normalized values related to the $y$ variable, defined in Eqs.~(\ref{eq:sigma0}) and (\ref{eq:sigmaInf}), and in order to obtain the original values for the variable $n$,
the values which appear in Eqs.~(\ref{eq:angrowth_n}), (\ref{eq:andamp_n}) and (\ref{eq:angrowth_smallk_n}) have to be multipled back by $k$ and  the corresponding sound speed,
therefore the actual values  do not depend on $k$. After replacing for $\sigma_T$ and $\sigma_\rho$ defined in Eq.~(\ref{eq:sigma0}), we obtain the 
following growth and damping rates for the weak coupling regime:
\begin{eqnarray}
\label{eq:angrowth_W}
g_{\rm W} =  \frac{\gamma-1}{2 \gamma} \rho_{\rm c0} \left[ \frac{2 \Lambda_0}{T_0} - \frac{\partial \Lambda_0}{\partial T} \right] \,,\\
\label{eq:andamp_W}
{\delta}_{\rm W} = \frac{\gamma-1}{4 \gamma} \rho_{\rm c0} \left[ \frac{2 \Lambda_0}{T_0} + (\gamma-1) \frac{\partial \Lambda_0}{\partial T} \right]\,,\\
\label{eq:angrowth_smallk_W}
g_{\rm W}^L=-\frac{\partial \Lambda_0}{\partial T} \frac{\gamma-1}{2} \rho_{\rm c0}\,.
\end{eqnarray}
The growth and the damping for the strong coupling regimes are the same as those obtained in the weak coupling regime,
multiplied by the factor $\frac{p_{\rm c0}}{p_{\rm 0}}$:
\begin{eqnarray}
\label{eq:angrowth_S}
g_{\rm S} = g_{\rm W}  \frac{p_{\rm c0}}{p_0}\,,\\
\label{eq:andamp_S}
{\delta}_{\rm S} = {\delta}_{\rm W} \frac{p_{\rm c0}}{p_0}\,,\\
\label{eq:angrowth_smallk_S}
g_{\rm S}^L = g_{\rm W}^L  \frac{p_{\rm c0}}{p_0}\,.
\end{eqnarray}
Therefore, when RC effects are added in a two-fluid model, the presence of the neutrals, 
sufficiently coupled to the charged particles by collisions,
inhibits the growth of the thermal mode and reduces the damping
on the sound waves by the same factor $\frac{p_{\rm c0}}{p_{\rm 0}}$.

The constant heating function instead of the time-varying heating function usually considered,
modifies slightly the isobaric and isentropic criterium, for having growth for the thermal mode and damping
on the sound waves, making them easier to be fulfilled. The growth rates and the damping rates are also slightly increased.
\end{itemize}

\subsection{Analysis of the dispersion relation}

\begin{table*}[]
    \centering
    \caption{Units.}
    \label{tab:units}
    \begin{tabular}{l l}
        \hline
        \hline
         Quantity & Unit \\[0.7ex]
        \hline
length & $U_L=10$~Mm\\
number density & $U_n=10^{9}$ /cm$^3$ \\
temperature & $U_T=10^6$ K \\
\hline
time & $U_t=110.1$ s \\
velocity & $U_v=90.85$ km/s \\
pressure & $U_p=0.138$ dyn/cm$^2$ \\
density & $U_\rho=1.67 \times 10^{-15}$ g/cm$^3$ \\
magnetic field & $U_B=1.32$ G \\
\hline 
    \end{tabular}
\end{table*}
\begin{figure*}
\FIG{
\includegraphics[width=14cm]{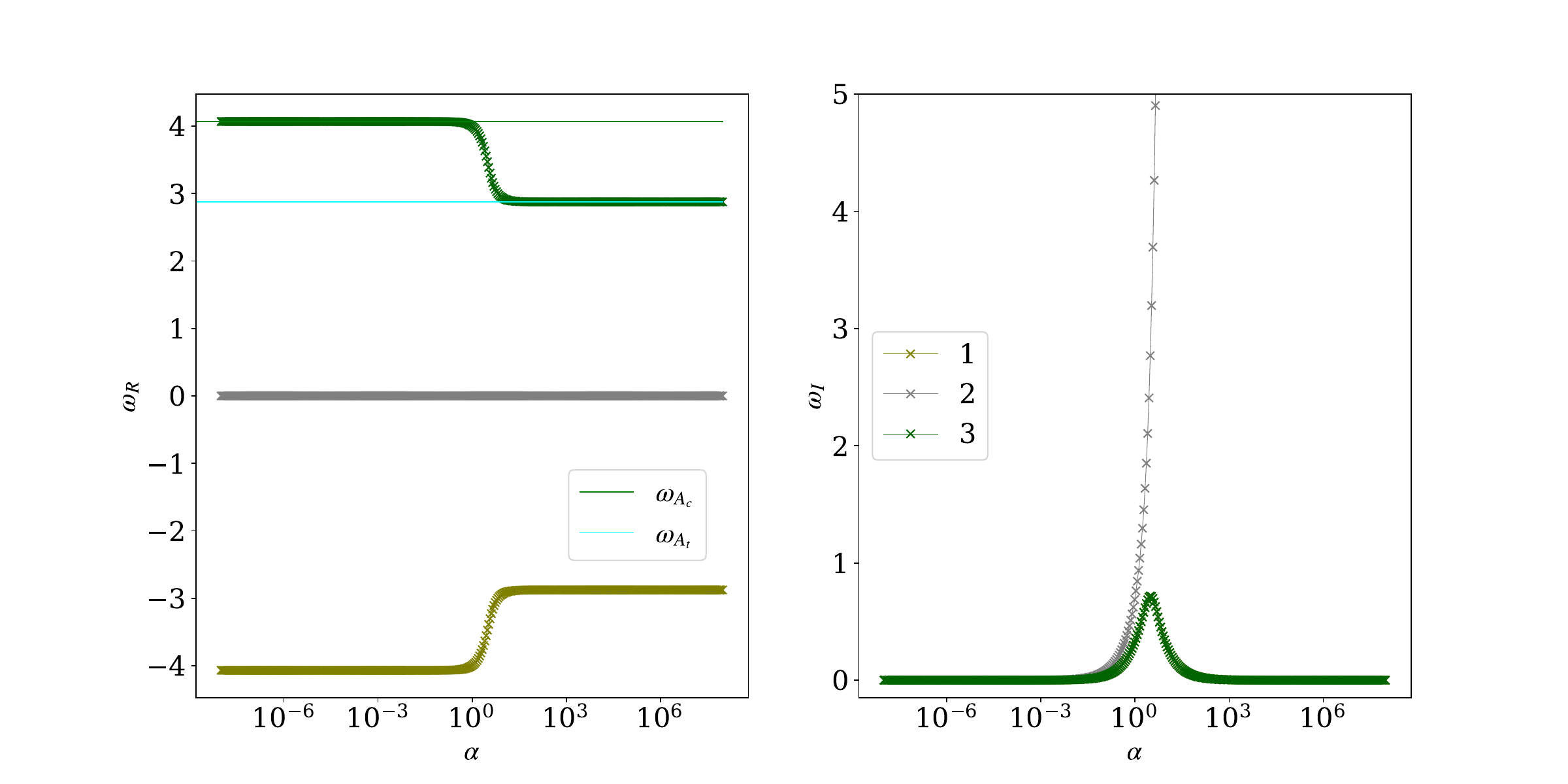}
}
\caption{The three solutions of the (cubic) Alfv\'en branch, Eq.~\ref{eq:alfbr}.
The Alfv\'en frequencies are shown { in the left panel by horizontal lines calculated as $\omega_{A_c} = v_{\rm Ac0} k_\parallel$, 
and $\omega_{A_t} = v_{\rm At0} k_\parallel$, with the charged fluid and total fluid Alfv\'en speeds,
defined in Eqs.~(\ref{eq:vac0}) and (\ref{eq:vat0}), respectively.}
We clearly recover the expected limit behavior in weak and strong coupling regimes.
}
\label{fig:2flalf}
\end{figure*}
\begin{figure*}
\FIG{
\includegraphics[width=14cm]{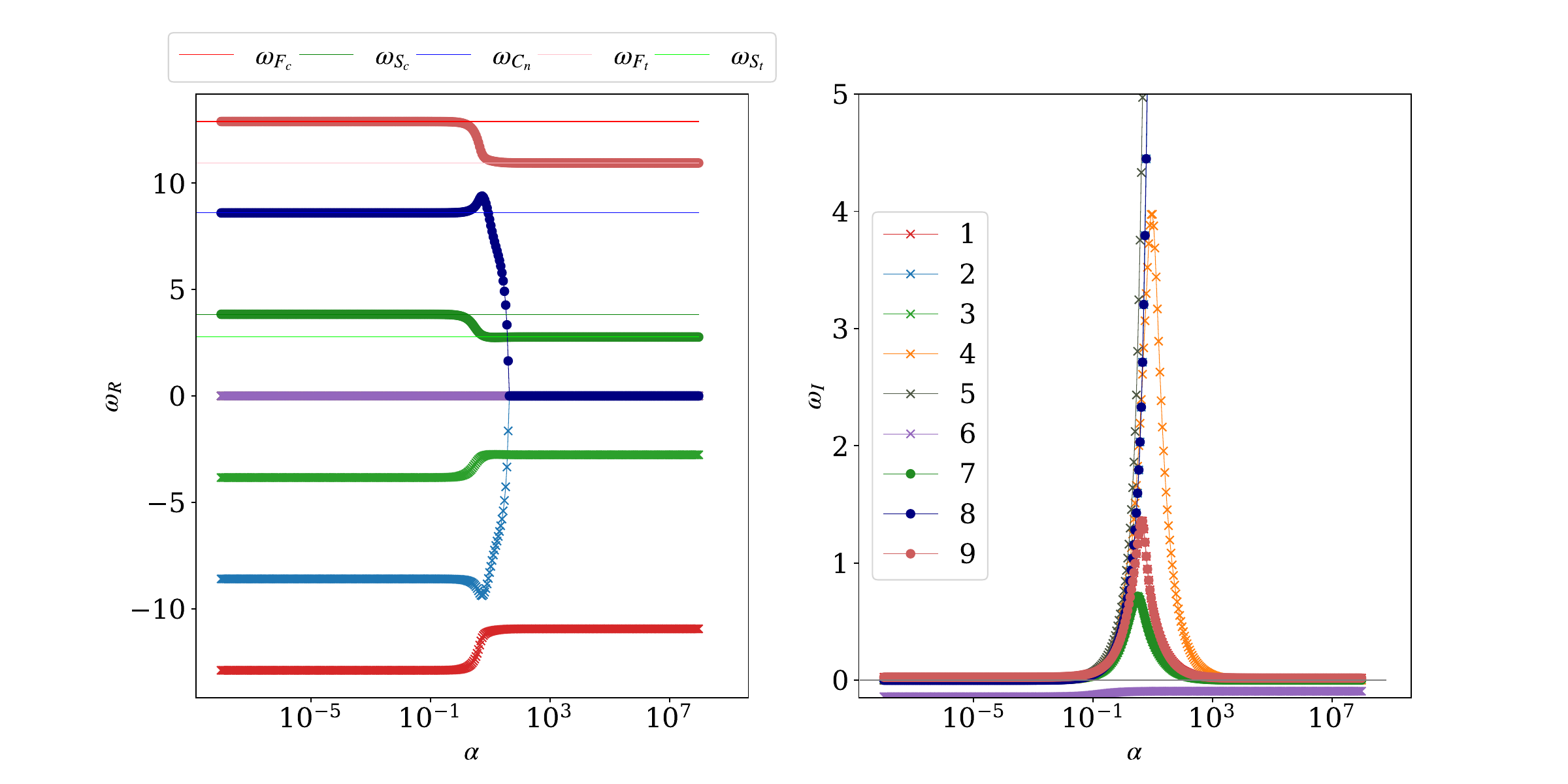}
}
\caption{The nine solutions of the 9th order term, containing 
six propagating modes:
\#1, \#9 fast (red); 
\#2, \#8 neutral sound (blue); \#3, \#7 slow (green, 'x' and 'o' for different directions); 
and three evanescent modes:
\#4 neutral entropy (orange,'x'), \#5 neutral shear (black, 'x'), \#6 entropy (violet, 'x'). 
{
The evanescent modes have $\omega_R=0$, therefore the three corresponding curves (orange, black and violet)
overlap in the left panel.
The frequencies  shown  in the left panel by horizontal lines are calculated as $\omega_{F_c} = f_{\rm c0} k$,
$\omega_{S_c} = s_{\rm c0} k$, $\omega_{C_n} = c_{\rm n0} k$, $\omega_{F_t} = f_{\rm t0} k$, $\omega_{S_t} = s_{\rm t0} k$,
with the characteristic speeds $f_{\rm c0}$, $s_{\rm c0}$, $c_{\rm n0}$, $f_{\rm t0}$, $s_{\rm t0}$  
defined in Eqs.~(\ref{eq:fc0}), (\ref{eq:ft0}), (\ref{eq:cn0}), (\ref{eq:ft0}), (\ref{eq:st0}), respectively.
In the right panel an horizontal black thin line $\omega_I=0$ is shown for clarity.}
Once more, limits at weak and strong coupling are fully understood.
}
\label{fig:2flcomp}
\end{figure*}

We will now consider a general case, when the propagation is not parallel to the magnetic field, and we will
solve numerically the dispersion relation. 
Solving the dispersion relation numerically does not give insight on the generality of the results as a purely analytic result,
however, it gives accurate results when analytic methods are not possible.
We can test the applicability of the analytical results obtained in limiting cases, presented previously, for this general case.

We will consider the same setup as we use later in the numerical simulations.
The setup is 2D, having the magnetic field angle to the direction of propagation $\theta=45^\circ$.
Therefore we have to consider the dispersion relation of 9th order which describes the compressible branch
in the general case, with no assumption about the angle $\theta$.
Eqs.~(\ref{eqs:lin3d}) are written in a non-dimensional form and we used the units shown in Table \ref{tab:units}.
We consider a total backgound density and pressure $\rho_0=1.375$ and  $p_0=2.318$, respectively. 
The background temperature is around $10^6$~K, the parameters being characteristic for coronal plasma. 
{ As previously mentioned, the inelastic processes are not taken into account in this part.}
The scale height corresponding to this temperature is about 154~Mm for the neutrals and twice for the charges,
therefore our domain of 10~Mm is much smaller than the scale height and the uniform medium is a good approximation.
We use the cooling curve ``ColganDM'' \citep[see][]{joris}, unless stated otherwise. 
The neutral and charged densities and pressures depend on the ionization fraction $\xi_i$ so that the temperatures of the two fluids are equal,
and for  Hydrogen only and charge neutrality:
\begin{eqnarray}
\label{eq:2flsplit}
\rho_{\rm c0} = \xi_i  \rho_0\,,\quad
\rho_{\rm n0} = (1 -\xi_i)  \rho_0\,,\nonumber\\
p_{\rm c0} = p_0 \frac{2 \xi_i}{1+\xi_i}\,,\quad
p_{\rm n0} = p_0 \frac{1-\xi_i}{1 +\xi_i}\,\quad
T_{\rm c0} =T_{\rm n0} = T_{\rm 0} = \frac{p_0}{\rho_0} \frac{1}{1 +\xi_i}\,.
\end{eqnarray}
We will show next the 12 solutions for $\omega$ to the general 3D system of Eqs.~(\ref{eqs:lin3d}).
For the parameters considered in our case, both conditions from  Eqs.~(\ref{eq:crit}) are satisfied, meaning that the thermal mode grows and the sound waves are damped.
In the low plasma $\beta$ regime, which is the case of the solar corona, 
the propagation speed of the slow  waves in the charged fluid is approximately equal to the sound speed of the neutrals for the angle $\theta=45^\circ$ considered. 
Therefore we considered for this plot a case with relatively high plasma $\beta\approx 2$  for a better visualization. 
We considered $\xi_i=0.5$,  $B_0=0.76$ and the mode number $n_w=1$.
{ The mode number $n_w$ is related to the wave number $k$ as $k= 2 \pi n_w/L$, where L is the length of the domain. 
Therefore  the wave number is $k=2\pi$, for $n_w=1$ and  a domain length equal to 1. 
}

Figure~\ref{fig:2flalf} shows the solutions of the Alfv\'en branch. 
The plots show the real and the imaginary part of the wave temporal frequency $\omega$ as a function of the
collisional parameter $\alpha$. The real part of the temporal frequency is related to the propagation speed and the imaginary part to changes of the amplitude, a negative value
means growth, while a positive value means damping.
There are two propagating modes ($\omega_R\ne 0$), being the Alfv\'en waves  propagating in both directions with a speed corresponding to the Alfv\'en speed calculated using charged fluid properties 
for small $\alpha$ and using the total fluid
properties for large $\alpha$. 
The Alfv\'en waves have maximum damping for intermediate coupling ($\omega_R\approx \alpha \rho_{\rm 0}$).
The third solution in this cubic branch is an evanescent wave ($\omega_R=0$), which is strongly damped for increasing $\alpha$. 
These waves were studied in detail by \cite{2013Soler}.

Figure~\ref{fig:2flcomp} shows the solutions of the 9th order dispersion relation corresponding to the compressible branch. The magneto-acoustic waves were studied by \cite{2013SolerB}.
We identify six solutions corresponding to propagating waves, in pairs for the two propagation directions 
and three evanescent modes.
The three pairs of propagating modes are the neutral sound, and the charges fast and slow waves.
For weak coupling ($\alpha \to 0$)   
the propagation speed of the fast and slow 
modes correspond to the charged fluid properties { ($f_{\rm c0}$ and $s_{\rm c0}$, from Eqs.~(\ref{eq:fc0}) and (\ref{eq:sc0}), respectively)} 
and become the propagation speed for the total fluid { ($f_{\rm t0}$ and $s_{\rm t0}$, defined in Eqs.~(\ref{eq:ft0}) and (\ref{eq:st0}), respectively)} as the coupling increases.
When the coupling increases, the neutral sound mode becomes damped (the imaginary part is positive and increases).
We can also observe that the damping of the Alfv\'en wave is similar to the damping of the slow wave (the slow wave has Alfv\'enic properties,
with motions almost perpendicular to the field lines, in high $\beta$ plasma regime). The Alfv\'en wave damping is smaller than the damping of the fast wave and the neutral wave.

The largest damping of the three evanescent modes (corresponding to solution \#5), increasing with increasing $\alpha$, is very   
similar to that found for the evanescent wave on the Alfv\'en branch (solution \#2 in Figure~\ref{fig:2flalf}). In order to understand if the dynamics of the mode
is dominated by the magnetic forces or the plasma pressure gradient
we can choose an eigenvalue on the curve and look at the eigenvector associated to it, which contains the amplitudes of the variables. 
We choose the eigenvalues corresponding to $\alpha=10^2$ (strong coupling) and $\alpha=10^{-6}$ (weak coupling). { The 
eigenvalues and eigenvectors for these two cases and for the two branches are shown in the Appendix \ref{sec:ev}}.
{ We can observe that for the mode corresponding to solution \#5, in the weak coupling case,
the perturbation is mostly in the velocity of the neutrals
 in  the direction perpendicular to $\mathbf{k}$, thus, this mode should be called the neutral shear wave.
The properties of this mode are indeed similar to those of the evanescent mode on the Alfv\'en branch.
In the strong coupling regime,  
both neutrals and charges oscillate in the direction perpendicular to $\mathbf{k}$ in high beta plasma regime (as shown in these Figures) or almost along the field lines 
in low beta plasma regime ($B_0=10$), keeping the total momentum constant. 
}

The mode corresponding to solution \#4 is mostly damped for intermediate coupling.
For the weak coupling regime the perturbation is mostly for the density of the neutrals, thus this mode is the neutral entropy mode.
In the strong coupling regime, there are perturbations in the densities and pressures of neutrals and charges,
 so that the total pressure is kept constant  and the temperature perturbation is the same in neutrals and charges.

The mode corresponding to solution \#6  is the only one which is growing ($\omega_I<0$).
In the weak coupling regime, the perturbation is mostly in the charged fluid density, 
therefore this is called the (charges) entropy mode or the thermal mode. 
When the cooling term is not included in the equations,  the thermal mode does not grow, having $\omega=0$.
In the strong coupling regime, the perturbation is mostly in the densities of neutrals and charges, 
with an amplitude ratio equal to the ratio of the background densities.
The perturbations are almost isobaric, similarly to the single-fluid case \citep{field}. 

The radiative cooling also produces damping on the slow and fast modes, but this is much smaller than the damping produced by collisions for intermediate coupling regime  and cannot be distinguished visually in the plots. 
For both solutions \#4 and \#6 the magnetic field is only slightly perturbed, creating even smaller perturbations in the velocity, aligned with the magnetic field,
as in the single-fluid case \citep{field}. 
As these perturbations in the magnetic field and velocity are orders of magnitude 
smaller than those in density we can conclude that the linear evolution of the modes corresponding to solutions \#4 and \#6 
is not determined by the magnetic field.
%
%
\begin{figure}
\FIG{
\includegraphics[width=8cm]{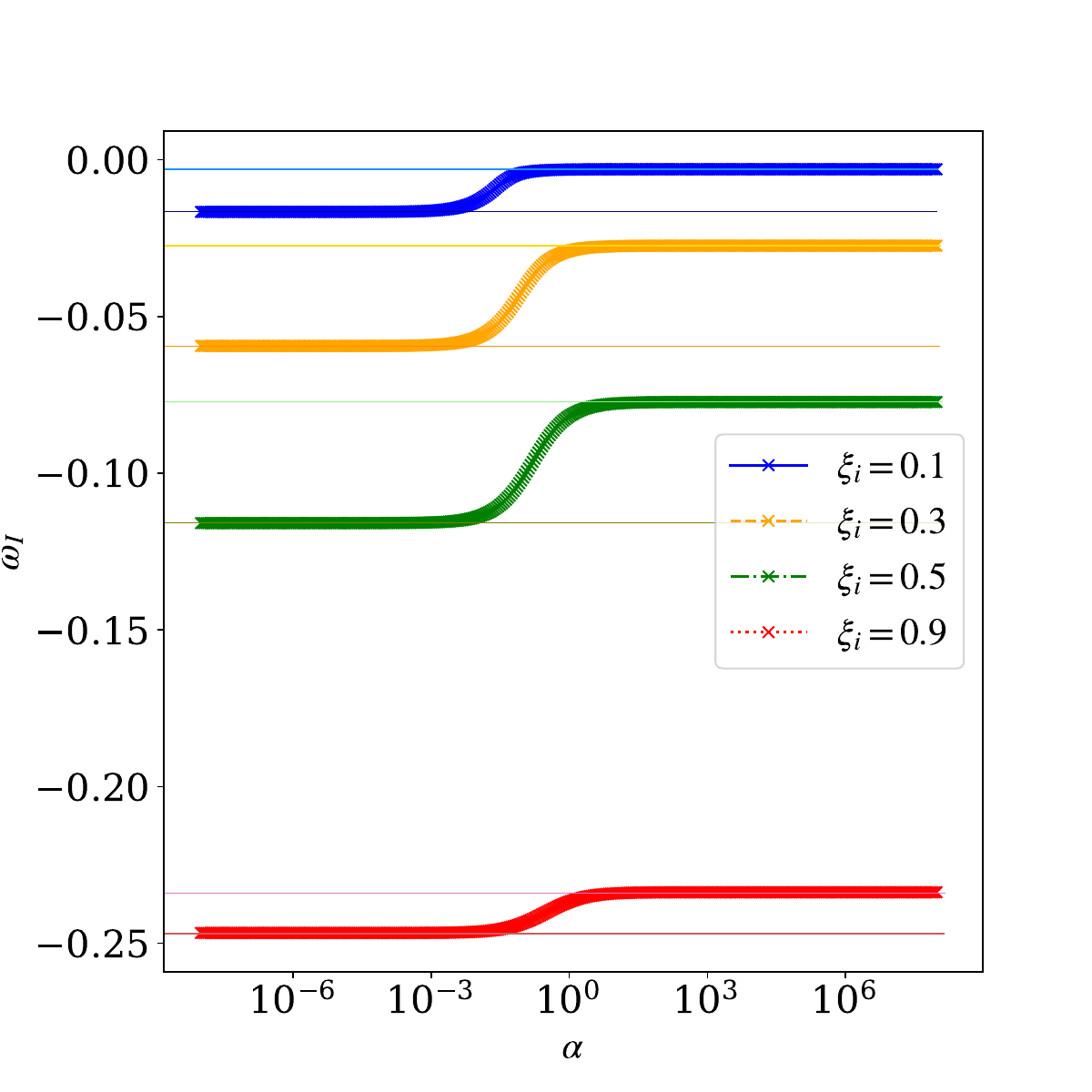}
}
\caption{Thermal mode growth as a function of the collisional parameter $\alpha$ for different ionization fractions.
The thin horizontal lines show the analytical growth rates obtained from Eq.~(\ref{eq:angrowth_W}) for $\alpha\to0$
and  Eq.~(\ref{eq:angrowth_S}) for $\alpha\to \infty$.
}
\label{fig:tifr}
\end{figure}
\begin{figure}
\FIG{
\includegraphics[width=8cm]{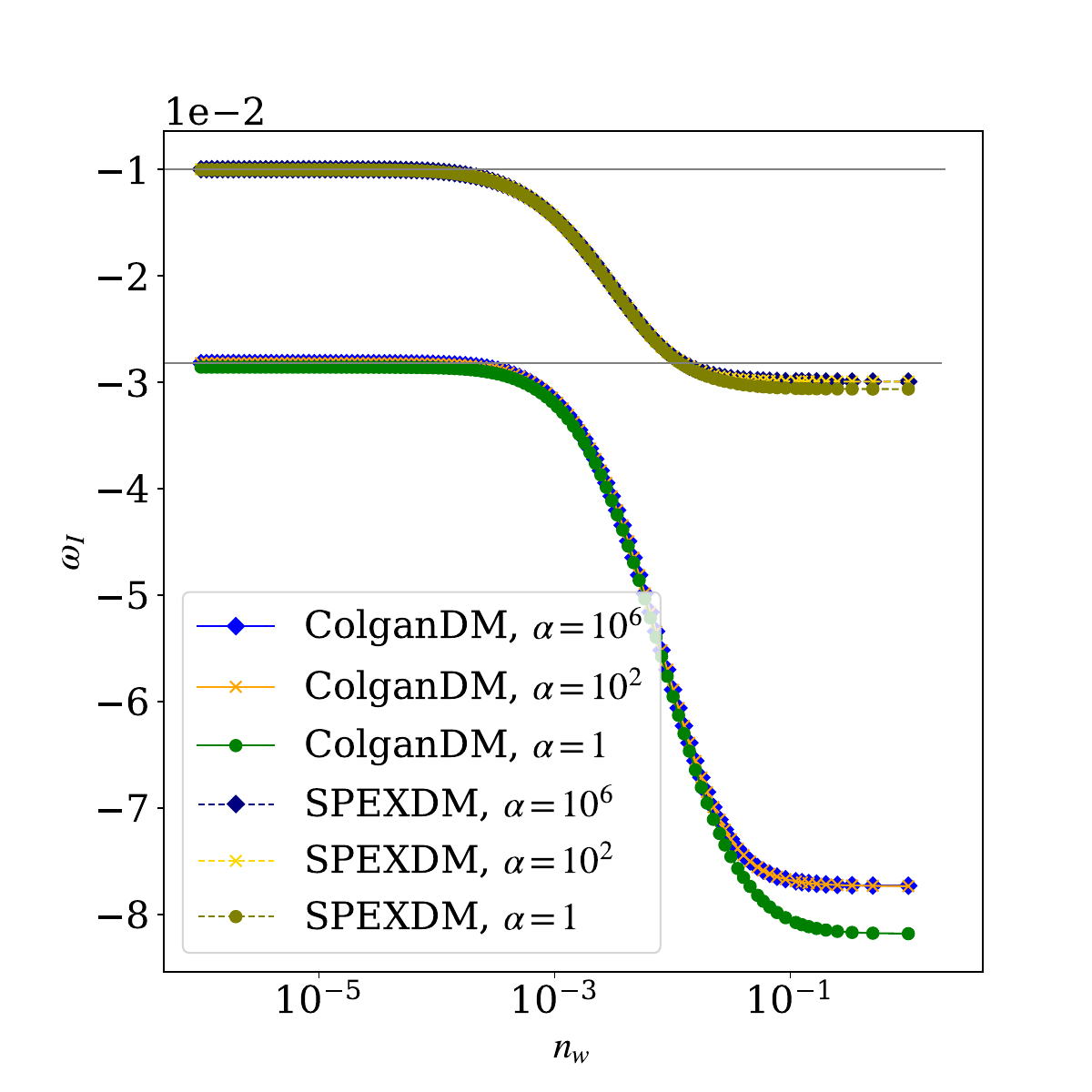}
}
\caption{
Growth of the thermal mode
as a function of the mode number $n_w=\frac{1}{m}$, with $m$ with 
integer values  from 1 to $10^6$. The fractional mode number corresponds
to scales larger than the length of the domain, having $k=2\pi n_w$.
We compare two cooling functions ``ColganDM'' (marked with ``x'') and ``SPEXDM'' (marked with ``o'')
and three values for the collisional parameter $\alpha$ indicated by three different colors.
The horizontal black lines show the analytical growth rates 
calculated in the high coupling regime for the $k\to 0$ limit, as 
from Eq.~(\ref{eq:angrowth_smallk_S}), for both curves. 
}
\label{fig:tik}
\end{figure}
\begin{figure*}
\FIG{
\includegraphics[width=7cm]{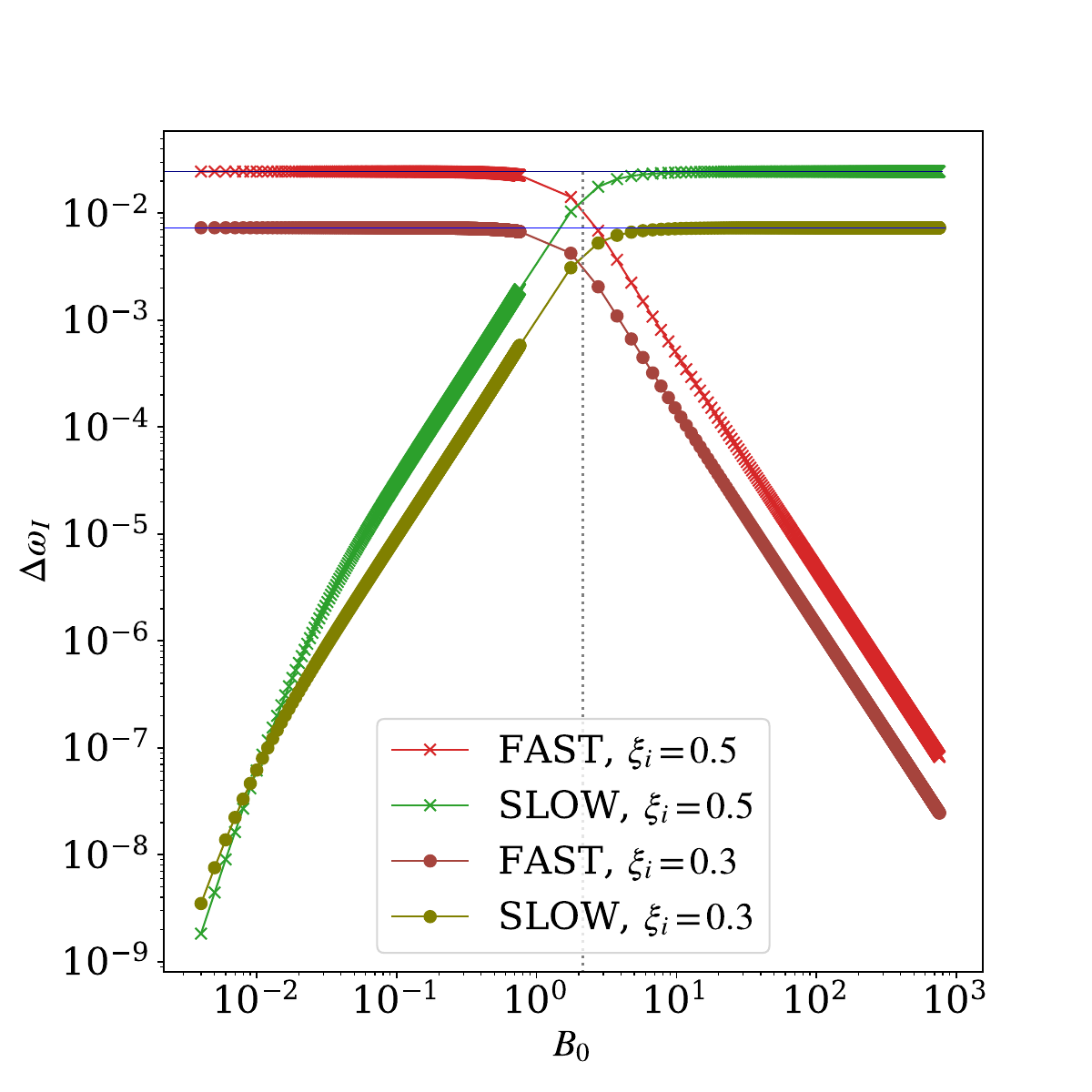}
\includegraphics[width=7cm]{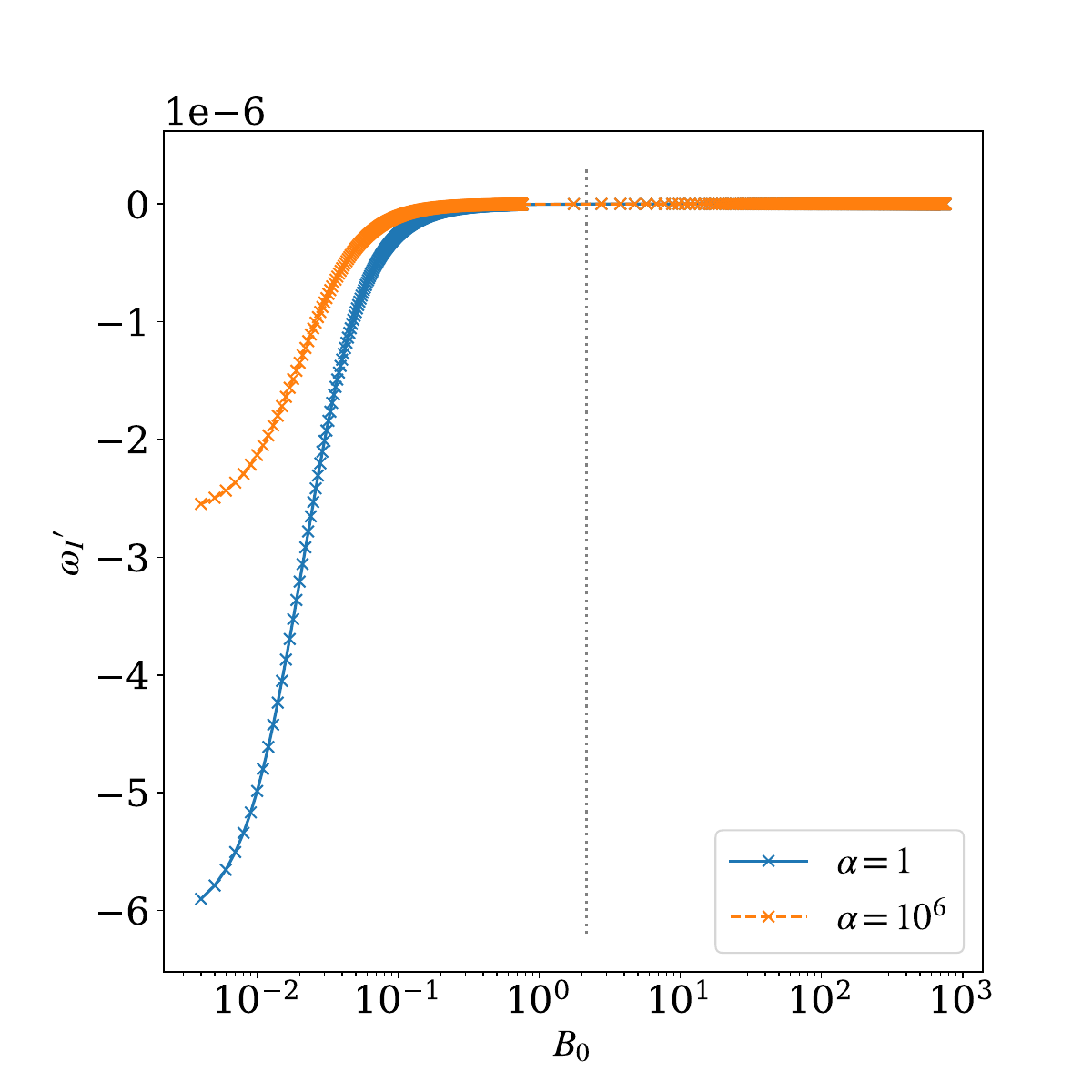}
}
\caption{
The dependence on the magnetic field magnitude.
Left: Damping due to radiative losses $\Delta \omega_I=\omega_I - (\omega_I)_{\rm noRC}$ for fast (green) and slow (red) modes,
for the strongly coupled case $\alpha=10^6$.
The horizontal blue lines show the damping of the 1D sound mode in high coupling regime as from  Eq.~(\ref{eq:andamp_S}). 
Right: Imaginary part of the thermal mode compared to the low beta limit growth:
$\omega_I\prime=\omega_I-(\omega_I)_{B_0=758}$. 
The low beta limit of the imaginary part $(\omega_I)_{B_0=758}$ was subtracted for a better visualization 
because it has values which are four orders of magnitude larger: 
$(\omega_I)_{B_0=758,\alpha=1} = -0.082$ and $(\omega_I)_{B_0=758,\alpha=10^6} = -0.077$. 
The ionization fraction considered was $\xi_i=0.5$ and the mode number $n=1$.
The dotted verical gray line  shows the equipartition value $\sqrt{\gamma p_0}$ in both panels.
}
\label{fig:damprc}
\end{figure*}
%

Figure~\ref{fig:tifr} shows the imaginary part of the thermal mode as a function of the collisional parameter $\alpha$ for several values of the ionization fraction.
The values of the background atmosphere depend on the ionization  fraction as in Eqs.~(\ref{eq:2flsplit}), 
therefore a higher ionization  fraction
means a higher density and lower temperature for the  charged fluid. For the cooling curve used here, the 
radiative cooling effects are larger for smaller temperatures around the equilibrium value. 
Therefore the growth rate is increased for higher ionization fraction because of two effects: lower temperature and higher density in the charged fluid.
The ratio between the growth of the thermal mode in the high coupling regime and the weak coupling regime depends on the ionization fraction only: 
$\frac{p_{\rm c0}}{p_{\rm 0}} = \frac{2 \xi_i}{1 + \xi_i}$.
The limits for weak and strong coupling, as defined in Eqs.~(\ref{eq:angrowth_W}) and (\ref{eq:angrowth_S}),
respectively, are shown 
by thin, horizontal lines. 
We can see a perfect agreement, therefore the single-fluid analysis
through the weak and strong coupling  limiting cases gives an almost complete description
of the thermal mode growth in the two-fluid model. Those limits are obtained for the high wavenumber $k$ regime, indicating that
the length of our domain is a small scale for the cooling effects. 

Figure~\ref{fig:tik} shows the imaginary part of the thermal mode as a function of the mode number $n_w$.
In the previous figure we considered $n_w=1$, and the limits obtained for high $k$ regime gave very good matching, indicating that the scale corresponding to the domain length is a small scale for the radiative cooling effects.
Therefore, we considered scales from $m=1$ up to $m=10^6$ larger than our domain,
having the fractional mode number $n_w=\frac{1}{m}$.
We compared two different cooling functions ``ColganDM'' and ``SPEXDM'' \citep[see also][]{joris} 
and three values of the collisional parameter $\alpha$.
We can observe that for both cooling functions the curves for different collisional parameters $\alpha$ converge to the same value
as $k\to 0$; the collisional mean free path becomes smaller than the inifinitely large scale associated to $k$, regardless the value of $\alpha$.
The growth rates for the ``SPEXDM'' curve are smaller than for the  ``ColganDM'' curve,  
being consistent with the results of the simulations presented in \cite{joris}.
The asymptotic value of the growth for the ``ColganDM'' cooling function at $n=1$ (corresponding to the high $k$ regime) is consistent to the values seen in Figure~\ref{fig:tifr}
for $\xi_i=0.5$ and the corresponding value of $\alpha$.
For the parameters considered here,  $\frac{\partial \Lambda_0}{\partial T}<0$, therefore we can calculate the
analytical growth rate for $k\to 0$ case using Eq.~(\ref{eq:angrowth_smallk_S}), corresponding
to the  high coupling regime, since 
$k\to 0$ is
the infinite coupling limit in the two-fluid model,  seen as well in the convergence of the curves for different $\alpha$ to the same limit when $k\to 0$. 
These analytical values are shown in the plot by horizontal 
thin black lines and we can observe a perfect agreement.

Figure~\ref{fig:damprc} shows the effect of radiative losses on the damping of slow and fast waves (left panel) 
and the imaginary part of the thermal mode (right panel) when we vary the magnitude of the magnetic field.
We can observe smaller damping  around the equipartition field,
where the waves have mixed sound and magnetic properties.
At the equipartition point, the fast mode transforms from having sound wave properties (isotropic, motion mostly along the direction of propagation
for high $\beta$ regime) to Alfv\'enic wave 
(motion almost perpendicular to the magnetic field) and the slow mode transforms from an Alfv\'enic wave to a sound wave (motion mostly along the field lines for 
low plasma $\beta$ regime).
Both of the modes are damped while having sound properties, the fast wave for fields below the equipartition layer and the slow wave for fields above 
and we can find perfect agreement
with the analytical damping  rates.

In the single-fluid assumption, the magnetic field has no effect
on the linear evolution of the thermal mode if $\mathbf{k}\parallel\mathbf{B}$.
When $\mathbf{k}\perp \mathbf{B}$,  
 the growth is inhibited because of the increase of magnetic pressure in the condensed regions 
because  the field is frozen into the plasma \citep{field}.
We can observe in the right panel of Figure~\ref{fig:damprc} that, indeed, the influence of the magnetic field to inhibit the growth is very small, the difference in the growth for a large change in the magnetic field is four orders of  
magnitude smaller than the values of the growth for both coupling regimes.

\section{Simulations}
\label{sec:sim}

In order to better compare the effect of radiative cooling in the two-fluid to the single fluid model in numerical simulations, 
we will consider the same setup and uniform background values as in \cite{joris,niels}, where a 
uniform periodic 2D box is perturbed by (damped) magnetosonic waves that ultimately give way to the thermal instability which causes a condensation to form. 

{
The 2D domain used in the simulations has length equal to 1 in both dimensions. 
We used periodic boundary conditions. 
The magnetic field is oriented by an angle of $45^\circ$,
with respect to the $x$-direction, $B_{\rm x0}=B_{\rm z0}=B_0/\sqrt{2}$. 
We use the same values used in the previous section for the background variables and  $k=2 \pi$.
The perturbation is initially a superposition of two single-fluid slow waves, with the same parameters,
but different propagation angles of $45^\circ$ and  $-45^\circ$, with respect to the magnetic field,  described by:

\begin{eqnarray}
\label{eqs:perts}
v_{\rm cz} = V_1 \text{cos}(k  z) + V_2 \text{cos}(k x)\,,\\
v_{\rm nz} = V_1 \text{cos}(k  z) + V_2 \text{cos}(k x)\,,\\
v_{\rm cx} = V_2 \text{cos}(k  z) + V_1 \text{cos}(k x)\,,\\
v_{\rm nx} = V_2 \text{cos}(k  z) + V_1 \text{cos}(k x)\,,\\
B_{\rm x1}=  B_1 \text{cos}(k  z)\,,\\ 
B_{\rm z1}=  B_1 \text{cos}(k  x)\,,\\ 
p_{\rm c1} = P_1 \frac{p_{\rm c0}}{p_0} \left( \text{cos}(k z) + \text{cos}(k x) \right)\,,\\ 
p_{\rm n1} = P_1 \frac{p_{\rm n0}}{p_0} \left( \text{cos}(k z) + \text{cos}(k x) \right)\,,\\ 
\rho_{\rm c1} = R_1 \frac{\rho_{\rm c0}}{\rho_0} \left( \text{cos}(k z) + \text{cos}(k x) \right)\,,\\ 
\rho_{\rm n1} = R_1 \frac{\rho_{\rm n0}}{\rho_0} \left( \text{cos}(k z) + \text{cos}(k x) \right)\,. 
\label{eqs:perte}
\end{eqnarray}

\noindent
We used the following quantities characteristic for the slow waves \citep{bookmhd}:
\begin{eqnarray}
\omega = k s_{\rm t0}\,,\\
V_1 = A\,,\quad V_2 = \frac{B_{\rm x0}^2  k^2}{B_{\rm x0}^2 k^2 - \omega^2 \rho_0} V_1\,,\\
B_1=\frac{1}{\omega}  B_{\rm x0} k ( V_1 - V_2 )\,,\\
P_1 =\frac{1}{\omega}  c_{\rm t0}^2 k \rho_0 V_1\,,\\
R_1 = \frac{1}{\omega} k \rho_0 V_1\,, 
\end{eqnarray}

\noindent
with $s_{\rm t0}$ the propagation speed for the total fluid,defined in Eq.~(\ref{eq:st0}).
We chose the amplitude $A=10^{-3}$. 
}

We used the same (physical) values for density $\rho_0$ and pressure $p_0$, split 
according to the specified ionization fraction $\xi_i$ 
into neutral and charged fluids, 
as described in Eqs.~(\ref{eq:2flsplit}),
in order to have the same temperature for neutrals and charges.
We used two different values for the magnetic field magnitude with physical values of 1~G and 10~G, and we also varied the 
collisional parameter $\alpha$ and the ionization fraction $\xi_i$, as shown in Table~\ref{tab:params}. 

\begin{table*}[]
    \centering
    \caption{Parameters of the simulations. }
    \label{tab:params}
    \begin{tabular}{l l l l | l l}
        \hline
        \hline
         Number & $\xi_i$ & $\alpha$ & $B_0$ & Plasma $\beta$ & Name \\[0.7ex]
        \hline
1 & 0.3 & $10^6$ & 0.76 & 2 & I1-$\alpha$3-B1							\\
2 & 0.5 & $10^6$ & 0.76& 2 & I2-$\alpha$3-B1							\\
3 & 0.5 & $10^6$ & 7.6& 0.02 & I2-$\alpha$3-B2							\\
4 & 0.5 & $1$ & 7.6 & 0.02 & I2-$\alpha$1-B2							\\
5 & 0.5 & $10^2$ & 7.6 & 0.02 & I2-$\alpha$2-B2							\\
6 & 0.9 & $10^2$ & 7.6 & 0.02 & I3-$\alpha$2-B2							\\
7 & 0.9 & $10^2$ & 0.76& 2 & I3-$\alpha$2-B1							\\
\hline 
    \end{tabular}
\end{table*}

\begin{figure}
\FIG{
\includegraphics[width=8cm]{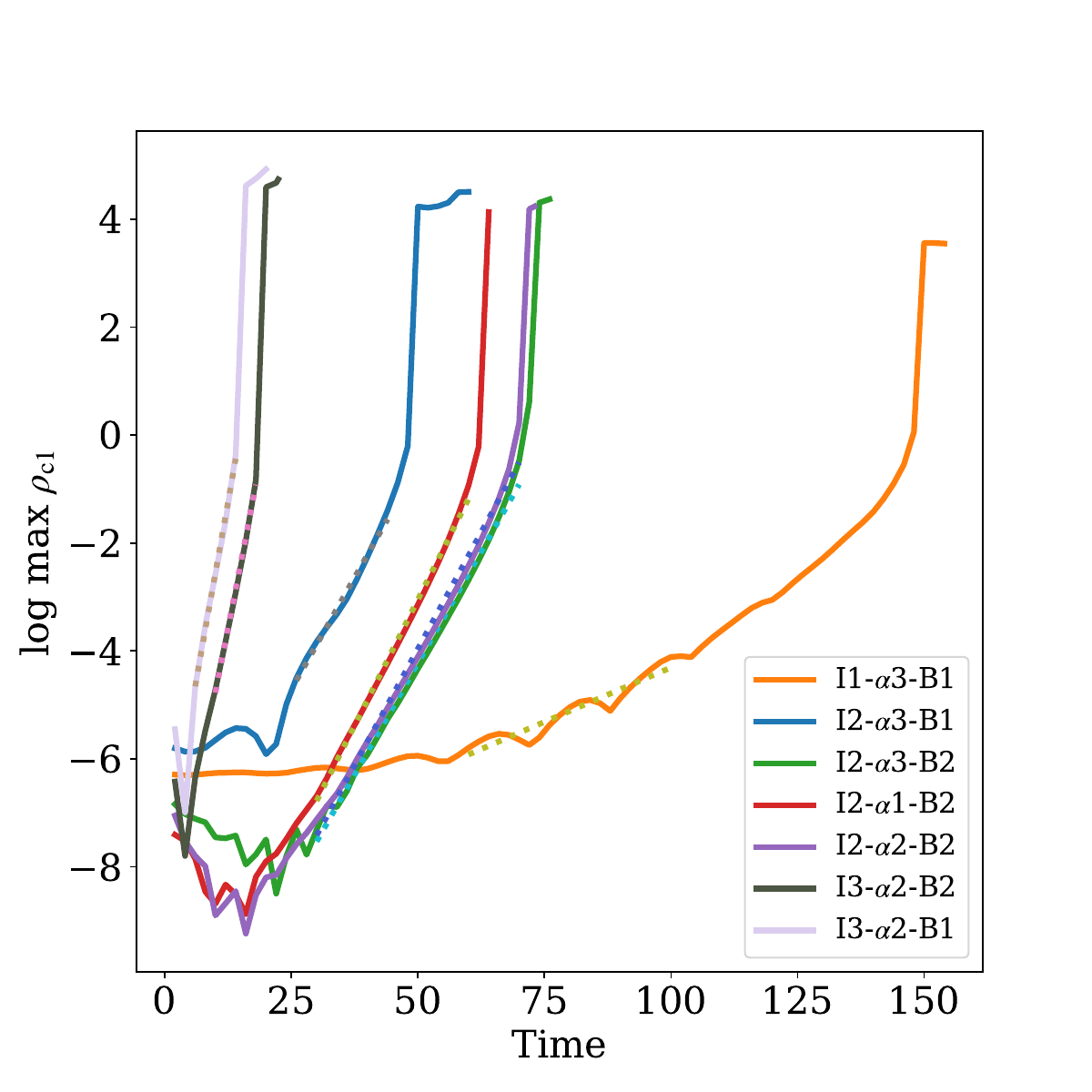}
}
\caption{
{ Time evolution of the logarithmic value of the maximum density of the charged fluid for the seven simulations shown in Table~\ref{tab:params}.
The linear growth rates can be calculated as the slopes of the curves during the linear phase.
The interval of time fitted for the linear phase is shown with dotted lines. The values of the fit and the errors as well as the analytical growth
rates are shown below for each of the simulations in Table~\ref{tab:lingr}.}
}
\label{fig:jg}
\end{figure}
\begin{table*}[]
    \centering
    \caption{The linear growth rates calculated from the simulations, with
    the linear phase indicated in Figure~\ref{fig:jg}, and
    the analytical growth rates, calculated numerically from the dispersion relation.}
    \label{tab:lingr}
    \begin{tabular}{l|l| l}
        \hline
        \hline
         Name & Growth rate sim. & Growth rate an. \\[0.7ex]
        \hline
 I1-$\alpha$3-B1					& 0.040 $\pm$ 0.003 & 0.027			\\
 I2-$\alpha$3-B1					& 0.166 $\pm$ 0.006 & 0.077			\\
 I2-$\alpha$3-B2				& 0.165 $\pm$ 0.003 & 0.077				\\
 I2-$\alpha$1-B2					& 0.187 $\pm$ 0.003 & 0.082			\\
 I2-$\alpha$2-B2				& 0.173 $\pm$ 0.004 & 0.077				\\
 I3-$\alpha$2-B2				& 0.482 $\pm$ 0.014 & 0.234				\\
 I3-$\alpha$2-B1					& 0.527 $\pm$ 0.014 & 0.234			\\
\hline 
    \end{tabular}
\end{table*}
\begin{figure*}
\FIG{
\includegraphics[width=14cm]{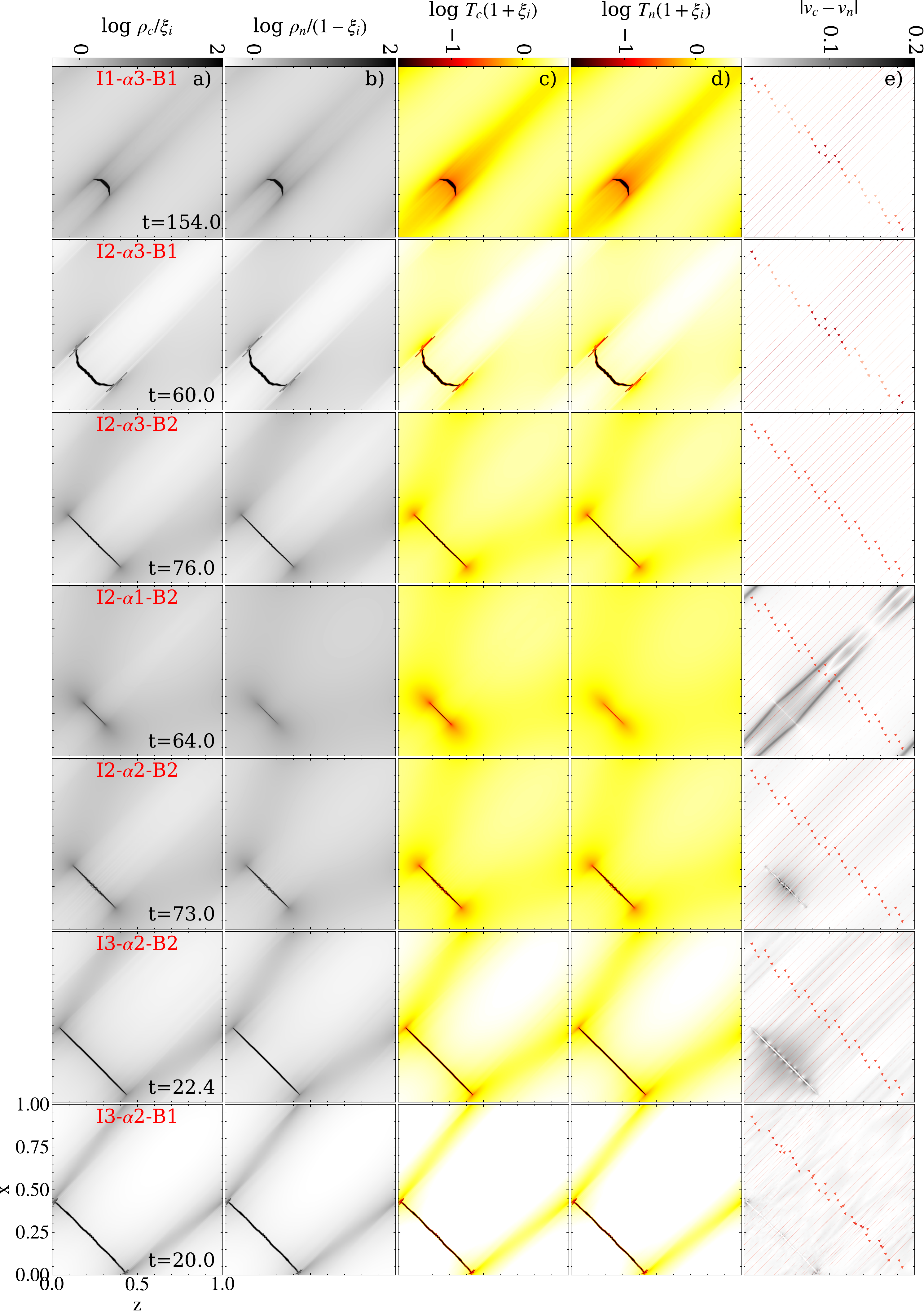}
}
\caption{
Last snapshots of the seven simulations shown in Table \ref{tab:params}
shown on the seven rows.
The columns correspond to different quantities:
density of charges, density of neutrals, temperature of charges, temperature of neutrals, magnitude of the decoupling in the velocity.
For a better visualization, in order to eliminate the effect of having different background neutral and charged 
densities and temperature for different ionization fractions, the quantities shown in the plots are normalized to have the background value equal to $\rho_0$ and the
temperature $p_0/\rho_0$.
The magnetic field lines are overplotted in the fifth column { with a red colormap, a darker color meaning a larger magnitude}.
}
\label{fig:jsnap}
\end{figure*}
Figure \ref{fig:jg} shows the evolution of the maximum of the perturbation in the density of charges as a function of time. 
{ The linear growth rates calculated from the simulations and the analytical ones, obtained
from solving numerically the 9th order dispersion relation are shown in Table~\ref{tab:lingr}.}
We can observe systematically for the seven simulations that the analytical results presented earlier are verified. 
The largest difference in the curves is given by changing the ionization fraction. 
Larger ionization fraction means larger growth, and this is consistent with Figure~\ref{fig:tifr}.
The difference for changing the ionization fraction is larger than other effects, therefore we observe
that the seven curves are grouped by this parameter. 
Within the groups, the growth rates are further ordered depending on the 
collisional parameter $\alpha$, smaller $\alpha$ meaning larger growth.
Changing the magnetic field magnitude has no significant effect on the growth rate.
There are small differences at the beginning because we do not excite exact
eigenmodes with the same amplitude.
We use the same perturbation in all the simulations \cite[the same as in ][]{joris}, representing single-fluid ideal MHD slow modes 
which are not eigenmodes of our system. 
The growth of the thermal mode can be seen after the slow waves are damped, therefore it is difficult to calculate
precisely the linear growth rates from the simulations. However, in the case when $\xi_i=0.3$, the growth  of the thermal mode 
is much smaller than  in the other cases; the damping rate is also smaller, but by a smaller factor (see Figures~\ref{fig:tifr} and \ref{fig:damprc}).
For this case, the calculated linear growth rate as the slope of the linear fit corresponds almost exactly to the analytical value. 
When the growth rates are larger, it is more difficult to separate the linear phase
to the nonlinear phase and the linear fit is not precise anymore, as we can 
see that the values obtained from the simulations deviate from the analytical growth rates 
for the larger ionization rates of 0.5 and 0.9, by a factor of almost two in the latter case.  
This is considered a very good agreement to the analytic predictions, 
given the way we initialize the simulations;
{ the ordering between the values is kept, and overall, qualitatively, the simulations growth rates follow the theory.}

Figure \ref{fig:jsnap} shows the last snapshots of densities (columns a) and b)), temperatures (columns c) and d)) and decoupling velocity (column e)) 
from the simulations shown in Table~\ref{tab:params}.
In the panel which shows the decoupling velocity we also show the magnetic field lines.
The evolution of the thermal instability is already in the nonlinear phase, when we can observe a high density
structure in the direction perpendicular to the magnetic field, in the bottom left part of the domain.
The similar location of the condensation for the seven simulations might be related to the decomposition of the initial perturbation (which contains only
one scale, $n_w=1$), in the nine eigenmodes of the system, 
and the values of the components of the eigenvector associated with the thermal mode, which give the initial (real)
amplitudes and phase shifts. It was mentioned previously that the very small velocity perturbed for the thermal mode is aligned with the magnetic field, even for very
small values of the magnetic field. 
By superposing two initial perturbations in the two dimensions { (see Eqs.~(\ref{eqs:perts})-(\ref{eqs:perte}})), 
these velocities have opposite signs making the structure stable \citep{joris}. 
In the nonlinear phase, the magnetic tension prevents the misalignment of the multiple scales 
formed later due to the nonlinear interaction. 
Smaller value for the magnetic field implies larger velocities as the magnetic field lines are easier to be compressed. This is because of the velocity perpendicular to the field created by the  gradient in the direction perpendicular to the field of the velocity parallel to the field.
This can be seen in the compressed field lines seen in the panels corresponding to the smaller field B1. In the panel I3-$\alpha$2-B1, we can observe  that the magnetic tension plays a role as well.

Larger ionization fraction produces larger structures as we can see by comparing the pairs I1-$\alpha$3-B1, I2-$\alpha$3-B1 and
I2-$\alpha$2-B2, I3-$\alpha$2-B2.
We can observe that for the simulations I1-$\alpha$3-B1, I2-$\alpha$3-B1 and
I3-$\alpha$2-B1 the condensation starts to be slightly misaligned to the perpendicular direction to the magnetic field.
The stronger  magnetic tension prevents the misalignment for the other cases.
We can also observe that the condensation is larger for larger ionization fraction, but this might be due to the fact
that the growth is larger.
Lower values of $\alpha$ makes the structures more diffusive and are also associated to larger values in the decoupling velocities.
The low temperature structures follow mostly the  high density structures  and are very similar for the neutrals and the charges.
In the temperature panels, we observe some regions of increased temperature at the end of the condensations, especially for smaller
values of $\alpha$, indicating the role of the frictional heating.


\subsection{Including ionization/recombination}


\begin{figure*}
\FIG{
\includegraphics[width=14cm]{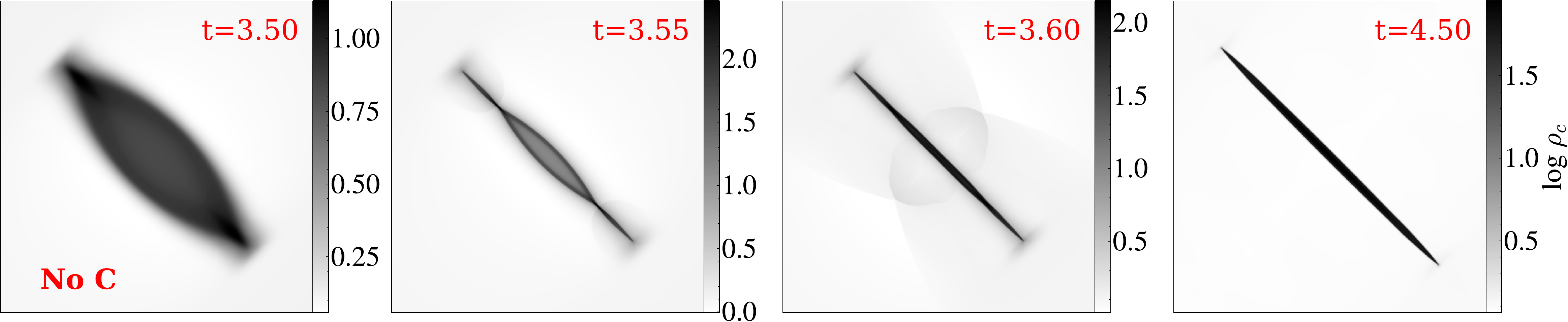}
\\ \rule[1ex]{14cm}{0.25pt} \\
\includegraphics[width=14cm]{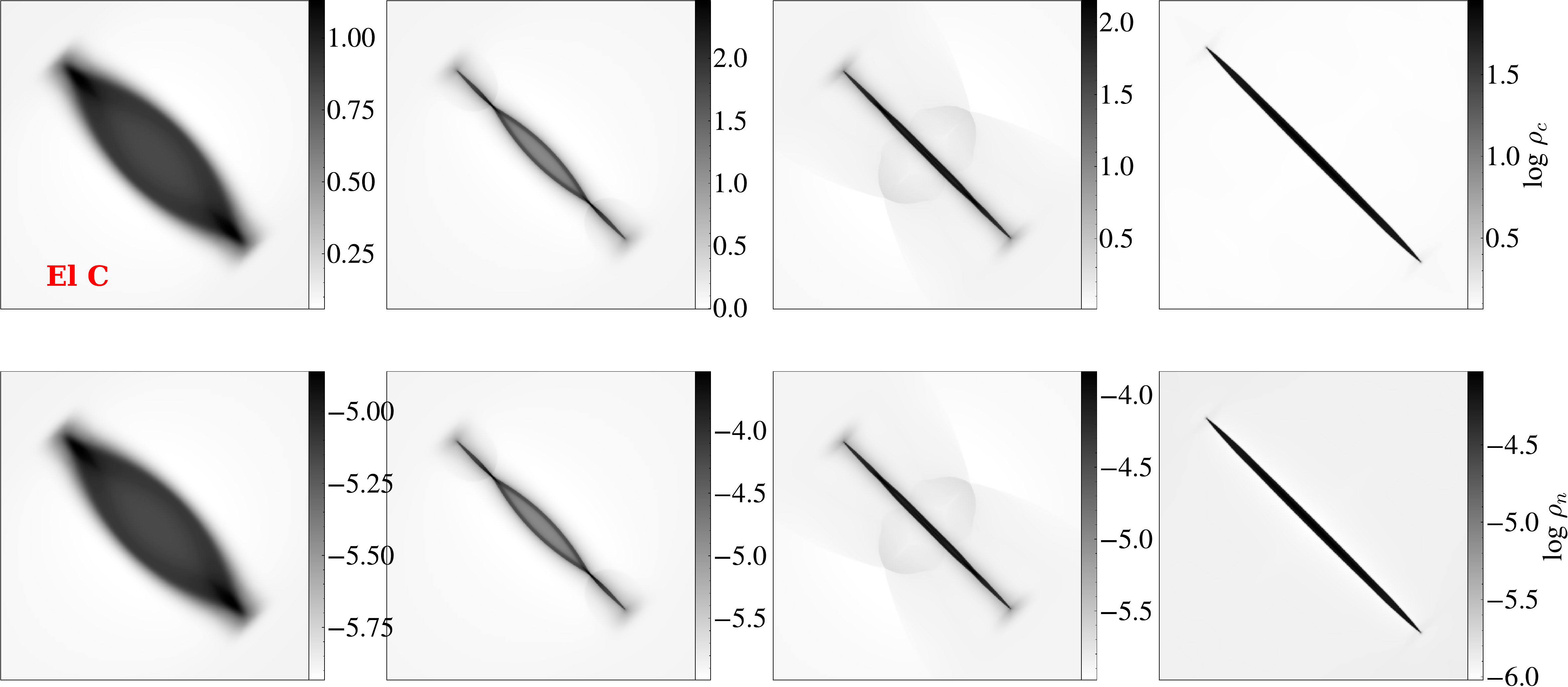}
\\ \rule[1ex]{14cm}{0.25pt} \\
\includegraphics[width=14cm]{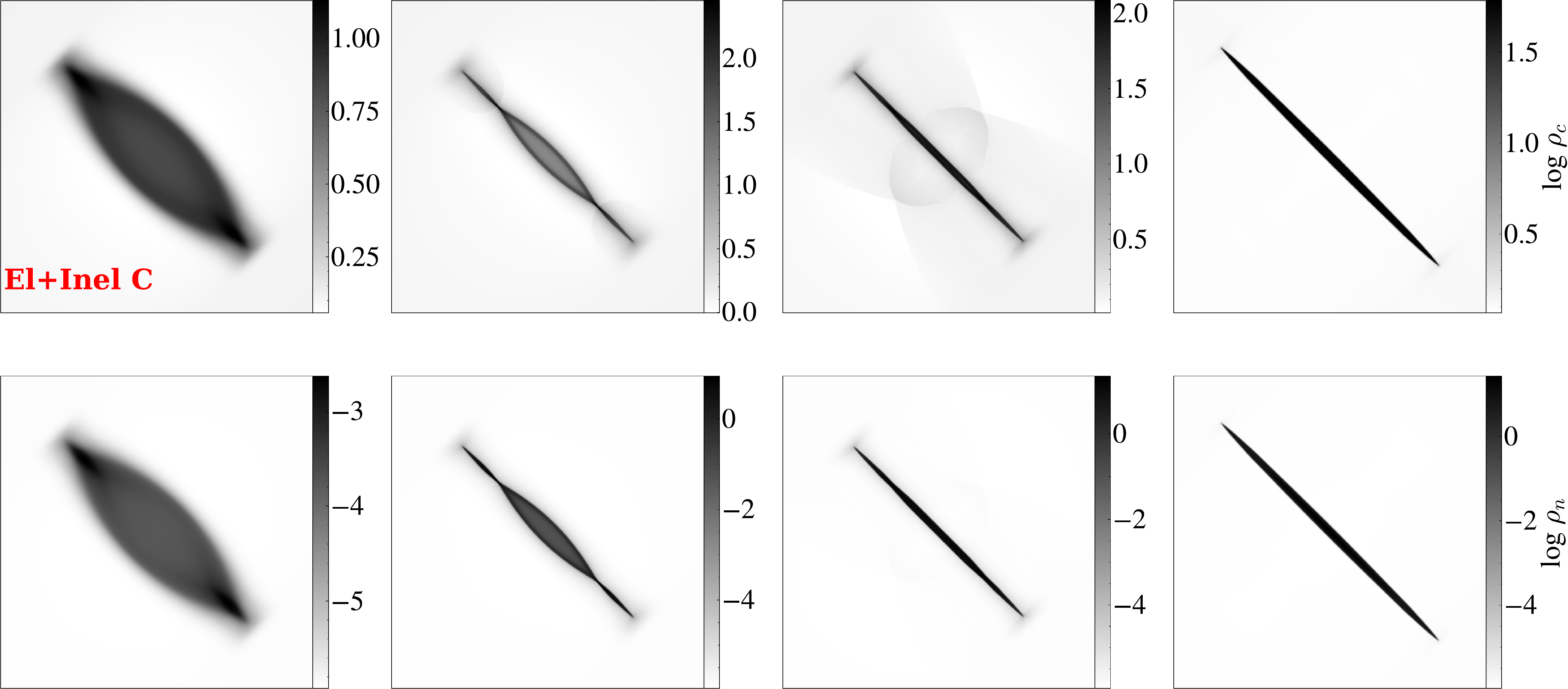}
}
\caption{
Time evolution of the densities  for ``No C'', ``El C''  and ``El+Inel C''.
For ``No C'' we show only the density of charges, because the neutrals do not evolve at all, since there is no coupling with the charges.
The movies of the three simulations are available online.
}
\label{fig:irsnap}
\end{figure*}
\begin{figure*}
\FIG{
\includegraphics[width=14cm]{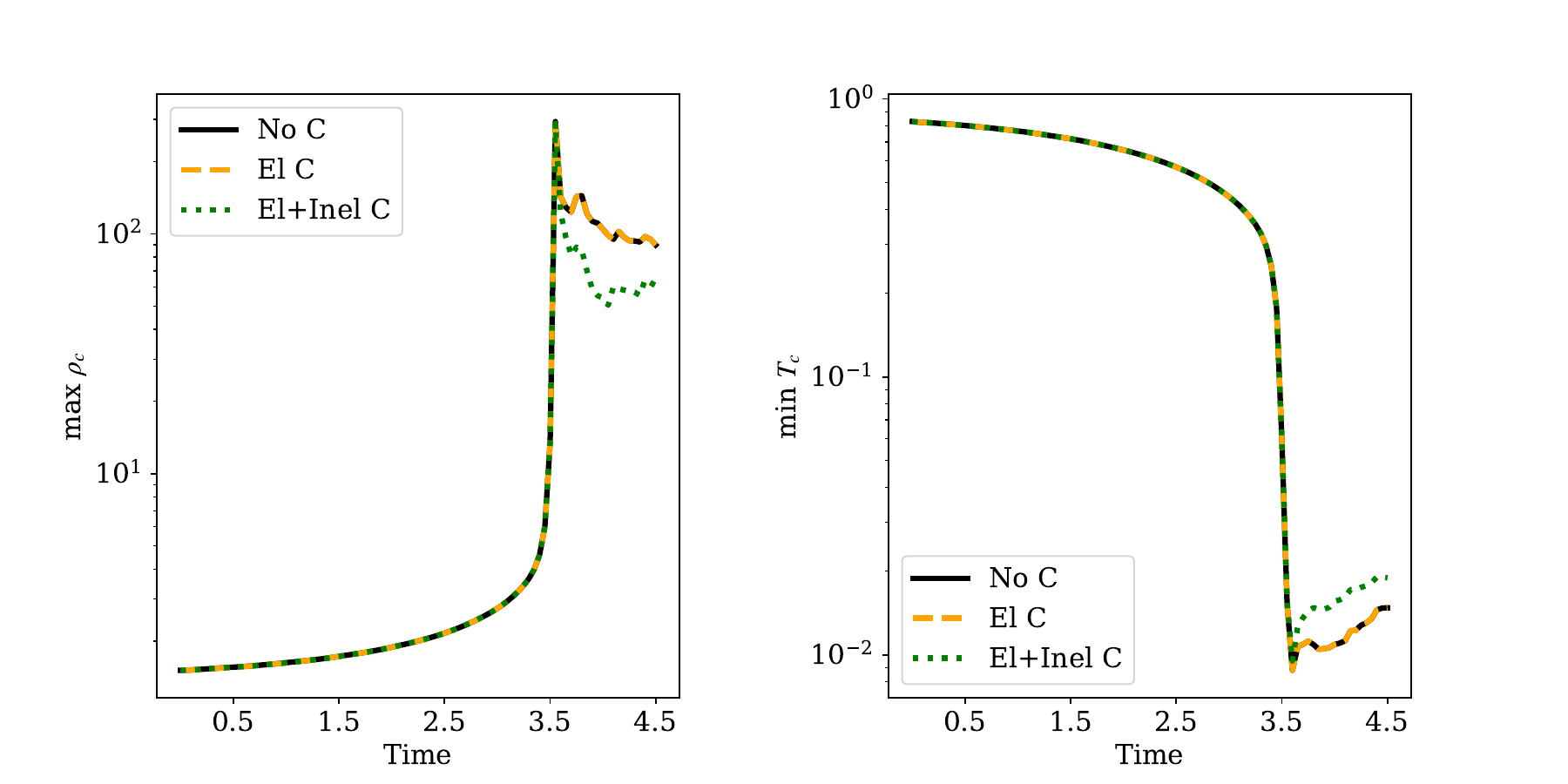}
}
\caption{
Evolution of the maximum density of charges as a function of time for the three simulations:
``No C'', ``El C''  and ``El+Inel C'', shown by the different curves as indicated in the legend.
}
\label{fig:irMaxDens}
\end{figure*}
\begin{figure}
\FIG{
\includegraphics[width=8cm]{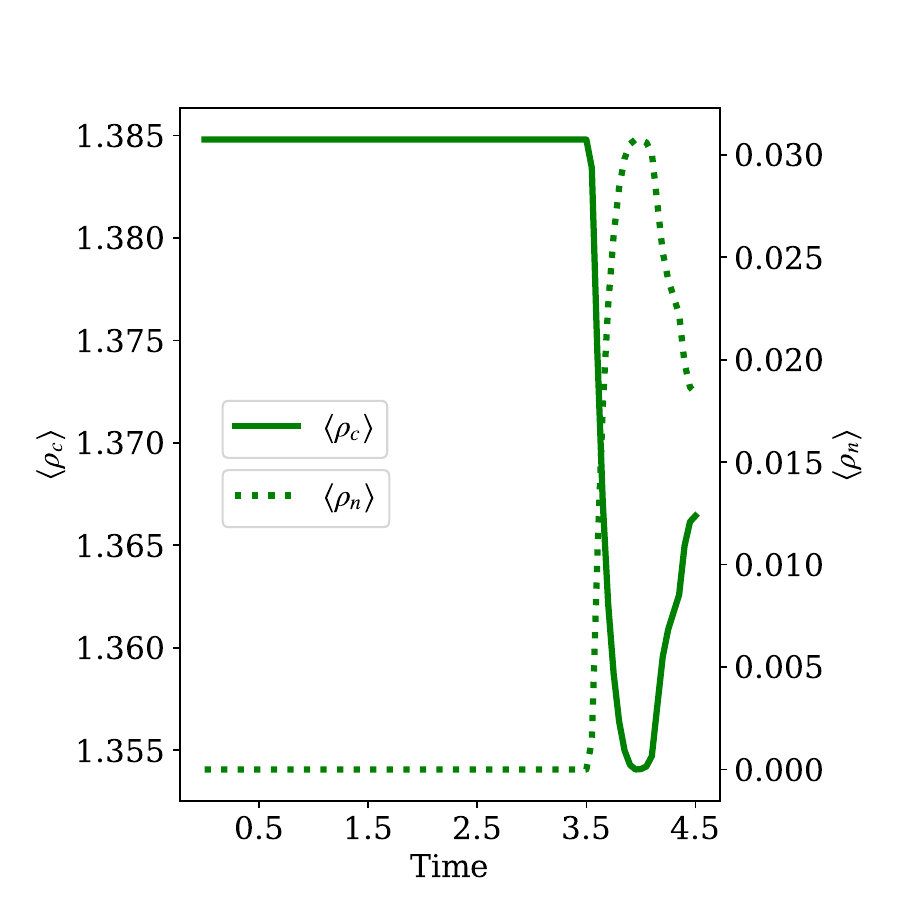}
}
\caption{
Evolution of the mean density of charges (solid line, left axis) 
and neutrals (dotted line, right axis) as a function of time in the simulation ``El+Inel C''. 
}
\label{fig:irMeanDens}
\end{figure}

In the simulations presented so far the ionization fraction $\xi_i$ was an input parameter, and we varied it in order to test several regimes.
However, we can calculate a realistic ionization fraction from the ionization recombination equilibrium condition for the chosen background,
described by $\rho_0$ and $p_0$.
{
For a fixed $p_0$ and $\rho_0$, he background temperature depends on the ionization fraction as shown in Eq.~(\ref{eq:2flsplit}):
\begin{equation}
\label{eq:getif1}
T_0=\frac{p_0}{\rho_0 (1+\xi_i)}\,.
\end{equation}
In order to have ionization/recombination equilibrium, the source term calculated for the background in the continuity equation 
must vanish.
After obtaining the equilibrium temperature in eV,
\begin{equation}
\label{eq:getif2}
\tilde{T} = T_0 U_T \frac{k_B}{e}\,,\quad
\end{equation}
where $U_T$ is the temperature unit, defined in Table~\ref{tab:units},   
and setting the continuity equations source term  $S_{\rm n0}=0$, with the definition given in 
Eq.~(\ref{eq:collts}) and ionization and recombination rates from Eqs.~(\ref{eq:sns})-(\ref{eq:sne}),
makes the ionization fraction a function of the temperature only,
\begin{equation}
\label{eq:getif3}
\xi_i = \frac{\sigma^{\rm ion}}{\sigma^{\rm ion} + \sigma^{\rm rec}} = f(\tilde{T})\,.
\end{equation}
Starting with an arbitrary ionization fraction and corresponding temperature as from Eq.~(\ref{eq:getif1}),
we iterate over recalculating the ionization fraction using Eqs.~(\ref{eq:getif2}) and (\ref{eq:getif3}),
and the temperature. This procedure converges fast, in two iterations if we start with $\xi_i=1$ and three iterations if
we start  with $\xi_i=0$.  
We obtain  the ionization fraction, 
corresponding to a very low fraction of neutrals, $1 - \xi_i=1.019\times10^{-6}$. 
}

The ionization/recombination and the radiative cooling processes are not independent, as part of the radiative losses are due to the recombination.
We used ionization/recombination rates and cooling curves parametrized as function of temperatures and densities only, calculated in external models.
These rates should be calculated consistently by the model,
but this is a very difficult task to put in practice. 

For the following nonlinear two-fluid simulations where we include ionization-recombination, we will change the perturbation to be similar to the thermal runaway setup shown in the demo simulations of our code MPI-AMRVAC 3.0, see \cite{amrvac}, adapted for the two-fluid model.  In practice, we set up a circular region of higher density and lower temperature in a uniform field and pressure. This will then evolve by thermal runaway/instability, where a condensation forms that again is oriented perpendicular to the ($45^\circ$ angle) magnetic field.

We will compare three simulations: an uncoupled case ``No C'' when there is no coupling between neutrals and charges,
``El C'' when we consider only elastic coupling and ``El+Inel C'' when we include ionization/recombination effects besides the elastic collisions. The value of the collisional parameter $\alpha$ is now calculated self-consistently, from the plasma parameters and has the 
non-dimensional value $\alpha\approx 10^3$, being almost constant and uniform during the simulation.

We show in Figure \ref{fig:irsnap} the time evolution of the densities 
for the three cases. Visually, the structures in both densities look very similar for the three simulations, 
except for the ``El+Inel C'' case, where we see a significant larger fraction of neutrals { by four orders of magnitude}.
After the compression stops, both neutrals and charges expand,
following a rebound shock wave, seen for both charges and neutrals in the ``El C'' case, as it can be estimated in the panels of the third column at time $t=3.6$. The shock wave cannot be seen in the neutral fluid for ``El+inel C'', being damped by the recombination processes. 

Figure~\ref{fig:irMaxDens} shows the time evolution of the maximum of the density of charges (left panel)
and the minimum temperature (right panel) of charges for the three simulations.
The maximum in density is reached at the same time as the minimum in the temperature  as it can be also visually estimated from Figure ~\ref{fig:irsnap} and has similar
values for the three simulations. 
{ The decrease in temperature triggers the recombination of the charges, this being the cause of the large increase in the density of neutrals.}

As the temperature grows the cooling becomes active again and the plasma condensates again.
There are several cycles of condensation and cooling followed by heating and expansion, seen 
as small oscillations in the density and temperature curves. 
There is an extra heating effect coming from the elastic interaction with the neutrals in the simulations ``El C'' and ``El+inel C'', which can be slightly seen in the minimum temperature evolution.
For the ``El+inel C'' case, after the first peak, the maximum charges density drops visibly compared to the other two cases, because of the ionization/recombination. 

Figure \ref{fig:irMeanDens} shows the time evolution of the mean density of neutrals and charges for ``El+inel C''.
We can observe a significant drop in the density of charges and a corresponding increase in the density of neutrals { by four orders of magnitude} because of the recombination.
This happens after the minimum peak in temperature is reached, when the recombination rates are highest. 
Afterwards, when  the temperature starts to increase again, the ionization effects become more important, and the density of charges starts to increase again.
These heating-cooling cycles  affect the ionization/recombination processes and  it can be seen in the small
oscillations of  the  mean densities.

\section{Conclusions}
In this paper we studied the radiative cooling effects in a two-fluid model, by adding the radiative losses in the charges energy equation.
In the analytical derivation we considered a 1D case, thus discarding the effects of the magnetic field, 
and the limiting cases of weak and strong coupling
which can be treated analytically, following the derivation in \cite{field}. 
We find that the stability criterium for the thermal mode and for the sound waves, the isobaric and isentropic criteria, 
respectively, as defined in \cite{field} remain unchanged.
For the parameters considered in this paper, characteristic to the solar corona,  radiative cooling produces growth of the thermal mode and damping of the sound waves.
We highlighted the difference between having a constant heating function instead of time-varying as considered in the analytic derivation of \cite{field}.
The isobaric and isentropic criterium, for growth for the thermal mode and damping of
sound waves, are slightly easier to fulfill and growth and damping rates are  increased when we use the constant heating function.

Radiative cooling has no effect on Alfv\'en waves and damps the waves with sound properties when the restoring force is the pressure gradient.
We observed how the single-fluid analysis in a 1D geometry, without the magnetic field 
gives a very good description of the radiative loss effects. 
The growth of the thermal mode and damping of the waves are 
{ recovered from the single-fluid analysis for weak and strong coupling.}
The growth of the thermal mode and the
damping of the sound waves are reduced by the presence of the neutrals in the strong
coupling regime by the same factor, which depends on the ionization fraction only, $\frac{2 \xi_i}{\xi_i + 1}$.

The analysis of the linear equations and the simulation results showed
that in our case the largest difference in the growth of the thermal mode  
is given by changing the ionization fraction, 
among changing other parameters, such as the collisional coupling regime or the magnetic field strength,
because it means a change in the background atmosphere.
Changing the  ionization fraction means changing the background atmosphere, { increasing the ionization fraction means}  decreasing the  temperature
and increasing the density of charges, therefore
increasing the growth of the thermal mode and the damping on the sound waves. 
We also obtained larger growth rates for smaller values of the collisional parameter $\alpha$. In the nonlinear phase the structures are larger for larger ionization fraction and more diffuse for smaller values of the collisional parameter $\alpha$. In this case, when $\alpha$ is smaller,  we observe an increase in the temperature at the end of the condensed structures, which might be due to the frictional heating.

The magnetic field magnitude is not relevant for the thermal mode linear growth. 
The increase in the growth when the field magnitude goes to zero is four orders of magnitude smaller 
than the value of the growth found in very low $\beta$ plasma regime.
The damping on the slow and fast waves in regimes where they have sound properties
coincides with the damping obtained analytically for 1D sound waves for the corresponding coupling regime. 
The slow waves are damped in the low plasma $\beta$ regime and the fast waves are damped in the high $\beta$ plasma regime.
The damping is slightly smaller around the equipartition value for the magnetic field magnitude when both waves have mixed sound and magnetic properties.
The magnetic field  might become more important for perpendicular propagation, however these effects are small compared to the effects
of the parallel conductivity \citep{field}, which was not taken into account in this study. In the nonlinear phase, the condensed structures align perpendicularly to the magnetic field lines, as seen in the simulations by \cite{joris,niels}.

The recombination of the charges when the temperature drops leads to an increase in the neutral density of 
four orders of magnitude. Including ionization-recombination introduced more significant
effects in the evolution of the thermal instability than the elastic collisions { as seen from Figures~\ref{fig:irsnap} and \ref{fig:irMaxDens}}.

\section{Appendix A: Collisional terms}
\label{sec:collt}

\begin{eqnarray}
\label{eq:collts}
S_{\rm n} = \rho_{\rm c} \Gamma^{\rm rec} - \rho_{\rm n}\Gamma^{\rm ion}\,,\\
\mathbf{R}_{\rm n} = \rho_{\rm c} \mathbf{v}_{\rm c} \Gamma^{\rm rec}  - \rho_{\rm n} \mathbf{v}_{\rm n} \Gamma^{\rm ion} + 
\rho_{\rm n} \rho_{\rm c} \alpha (\mathbf{v}_{\rm c} - \mathbf{v}_{\rm n})\,,\\
M_{\rm n} = \frac{1}{2} \rho_{\rm c} v_{\rm c}^2 \Gamma^{\rm rec}  
- \frac{1}{2}\rho_{\rm n} v_{\rm n}^2 \Gamma^{\rm ion} 
+  \frac{1}{2} ({v_{\rm c}}^2 - {v_{\rm n}}^2) \rho_{\rm n} \rho_{\rm c} \alpha
\nonumber\\
+ \frac{1}{\gamma-1} 
\left ( \rho_{\rm c} T_{\rm c} \Gamma^{\rm rec} - \rho_{\rm n} T_{\rm n} \Gamma^{\rm ion} \right) 
 \nonumber\\
+\frac{1}{\gamma-1} (T_{\rm c} - T_{\rm n})\rho_{\rm n} \rho_{\rm c} \alpha\,.
\label{eq:collte}
\end{eqnarray}

Expressions for $\Gamma^{\rm ion}$ and $\Gamma^{\rm rec} $ as functions of $n_e$ and $T_e$ are given in \cite{1997Voronov} and \cite{2003Smirnov};
are those described by Eqs.~(A5) and (A4), respectively in \cite{paper2f}, 
the same as Eqs. (A.2), (A.1) from \cite{beatrice1}:

\begin{eqnarray} 
\label{eq:sns}
\Gamma^{\rm rec}  \approx n_e \sigma_{\rm rec} \,\,\,\, {\rm s^{-1}} \quad
\Gamma^{\rm ion}  \approx n_e \sigma_{\rm ion} \,\,\,\,  {\rm s^{-1}}\,,\\
\sigma^{\rm ion}=
\frac{A}{X + \frac{E_{\rm ion}}{\tilde{T}}} \left(\frac{E_{\rm ion}}{\tilde{T}}
\right)^K \text{exp}\left( \frac{E_{\rm ion}}{\tilde{T}} \right)\,,\quad
\sigma^{\rm rec}=\frac{1}{\sqrt{\tilde{T}}} (2.6 \times 10^{-19})\,,\\
\label{eq:sne}
\end{eqnarray}
$\tilde{T}$ is the temperature in eV,
$A = 2.91\times10^{-14}$ m$^3$/s,
K = 0.39,
X = 0.232,
$E_{\rm ion}$ = 13.6  eV.  

\section{Appendix B: Eigenvectors and eigenvalues}
\label{sec:ev}

Below we show the numerical values of the eigenvalues and eigenvectors
for the case $\xi_i=0.5$, $k=1$.
For a better readability, the real and imaginary parts with absolute values smaller than $2^{-23} = 1.19e-7$  (the float machine precision) are set to 0.
Each row of the eigenvectors matrix $V$ correspond to the eigenvalues shown in the vector $\Lambda$. 
The eigenvectors contains the amplitudes of:
$v_{\rm cy}$,  $v_{\rm ny}$, $B_{\rm y1}$, for the Alfv\'en branch and 
$p_{\rm c1}$, $p_{\rm n1}$, $v_{\rm cx}$, $v_{\rm nx}$ , $v_{\rm cz}$,$v_{\rm nz}$, $\rho_{\rm c1}$, $\rho_{\rm n1}$, $B_{\rm z1}$
for the compressible branch. 
All the calculations are done for $B_0=1$ (the case shown in Figures~\ref{fig:2flalf} and \ref{fig:2flcomp}), except for the coupled case, where we also calculated the case $B_0=10$ 
for the compressible branch, as indicated below.

\noindent
{\bf High beta regime ($B_0=1$)}

\subsection{$\alpha=10^{-6}$ (uncoupled case)}

\subsubsection{Alfven branch}
\label{sec:evAlfUnc}

\begin{equation*}
\Lambda=
 \begin{bmatrix}
 -4.07+3.44e-7 i \\ 
 6.88e-7 i \\ 
 4.07+3.44e-7 i\\
\end{bmatrix}
\end{equation*}

\begin{equation*}
\hspace*{-2.5cm}
V = \begin{bmatrix}
 0.77& 1.30e-7 i& 0.64\\
 0& 1 & -1.40e-7 i\\
 0.77& -1.30e-7 i& -0.64\\
 \end{bmatrix}
\end{equation*}

\subsubsection{Compressible branch}
\label{sec:evCompUnc}

\begin{equation*}
\Lambda=
 \begin{bmatrix}
 -12.89+0.03 i\\ 
 -8.60+4.81e-7 i\\
 -3.84+0.0045 i \\ 
 4.13e-7 i \\ 
 6.88e-7 i\\ 
 -0.12 i\\ 
3.84+0.0045 i \\ 
8.60+4.81e-7 i \\ 
12.89+0.03 i\\
 \end{bmatrix}
\end{equation*}

{\fontsize{6}{8}
\begin{equation*}
\hspace*{-2.5cm}
V = 
 \begin{bmatrix}
 0.75& 0& -0.60-0.0019 i& 0& 0.07+0.00058 i& 0& 0.20+0.0011 i& 0& 0.17+0.0011 i\\
 0& -0.64& 0& 0.68& 0& 0& 0& -0.34& 0\\
 -0.34-0.00024 i& 0& 0.08+0.0015 i& 0& 0.73& 1.32e-7 i& -0.09-0.0018 i& 0& 0.57-0.00067 i\\
 0& 0& 0& 0& 0& 0& -1.78e-6& 1 & 0\\
 0& 0& 0& 0& 0& 1& 0& 0& -1.40e-7 i\\
 -0.00068& 0& -0.03 i& 0& -0.03 i& -1.59e-7 i& 1 & 3.56e-6& 0.00063\\
 0.34-0.00024 i& 0& 0.08-0.0015 i& 0& 0.73& -1.32e-7 i& 0.09-0.0018 i& 0& -0.57-0.00067 i\\
 0& 0.64& 0& 0.68& 0& 0& 0& 0.34& 0\\
 0.75& 0& 0.60-0.0019 i& 0& -0.07+0.00058 i& 0& 0.20-0.0011 i& 0& 0.17-0.0011 i\\
 \end{bmatrix}
\end{equation*}
}

\subsection{$\alpha=10^{2}$ (coupled case)}

\subsubsection{Alfven branch}
\label{sec:evAlfCou}

\begin{equation*}
\Lambda=
 \begin{bmatrix}
-2.88+0.03 i \\ 
137.45 i \\ 
2.88+0.03 i
 \end{bmatrix}
\end{equation*}

\begin{equation*}
\hspace*{-2.5cm}
V = 
 \begin{bmatrix}
 0.54-5.69e-03 i& 0.54+0.02 i& 0.64\\
 -0.71& 0.71& -0.02 i\\
 -0.54-5.69e-03 i& -0.54+0.02 i& 0.64\\
\end{bmatrix}
\end{equation*}

\subsubsection{Compressible branch}
\label{sec:evCompCou}

\begin{equation*}
\Lambda=
 \begin{bmatrix}
-10.95+0.09 i \\ 
104.34 i \\ 
-2.77+0.04 i \\ 
0.43 i \\  
137.45 i \\ 
-0.08 i \\ 
2.77+0.04 i \\ 
135.71 i \\ 
10.95+0.09 i\\
\end{bmatrix}
\end{equation*}

{\fontsize{1}{3}

\begin{equation*}
\hspace*{-3.65cm}
V =  
 \begin{bmatrix}
 0.65& 0.33+0.01 i& -0.44+0.0051 i& -0.44-0.02 i& 0.03-0.0026 i& 0.03+0.0026 i& 0.17-0.0006 i& 0.17+0.0099 i& 0.15-0.0012 i\\
 0.69& -0.67& 0.19 i& -0.19 i& -0.00031 i& 0.00059 i& 0.0078& -0.0078& 0.0061\\
 -0.23-0.01 i& -0.11+0.02 i& 0.04+0.0044 i& 0.04-0.0091 i& 0.53-0.0028 i& 0.53+0.02 i& -0.06-0.0078 i& -0.06+0.01 i& 0.60\\
 -0.57& 0.57& -0.02 i& 0.05 i& -0.02 i& -0.02 i& -0.22& 0.54& -0.0039\\
 0.0054 i& -0.0039 i& -0.03& 0.03& -0.71& 0.71& 0.00081 i& -0.00079 i& -0.02 i\\
 -0.00047& 0.000039& -0.01 i& -0.01 i& -0.01 i& -0.01 i& 0.71& 0.71& 0.0004\\
 -0.23+0.01 i& -0.11-0.02 i& -0.04+0.0044 i& -0.04-0.0091 i& -0.53-0.0028 i& -0.53+0.02 i& -0.06+0.0078 i& -0.06-0.01 i& 0.60\\
 0.15 i& -0.11 i& -0.69& 0.70& 0.02& -0.02& 0.02 i& -0.02 i& 0.02 i\\
 0.65& 0.33-0.01 i& 0.44+0.0051 i& 0.44-0.02 i& -0.03-0.0026 i& -0.03+0.0026 i& 0.17+0.0006 i& 0.17-0.0099 i& 0.15+0.0012 i\\
\end{bmatrix}
\end{equation*}

}

\noindent
{\bf Low beta regime ($B_0=10$)}

\begin{equation*}
\Lambda=
 \begin{bmatrix}
 -42.81+5.9 i\\ 
104.80 i\\ 
 -7.33+0.14 i\\ 
 0.43 i\\
 136.72 i\\
-0.08 i\\ 
7.33+0.14 i\\ 
124.15 i\\ 
42.81+5.9 i\\
\end{bmatrix}
\end{equation*}

{\fontsize{1}{3}

\begin{equation}
\hspace*{-3.5cm}
V = 
\begin{bmatrix}
0.14+0.0084 i& 0.07+0.03 i& -0.41+0.05 i& -0.34-0.16 i& 0.38-0.06 i& 0.32+0.15 i& 0.04+0.00037 i& 0.03+0.02 i& 0.62\\          
0.66& -0.65& 0.23 i& -0.27 i& -0.05 i& 0.09 i& 0.0094& -0.01& 0.09\\                                                           
0.66& 0.33-0.0080 i& -0.30+0.0021 i& -0.30+0.01 i& -0.32+0.0019 i& -0.32-0.03 i& 0.18+0.0021 i& 0.18-0.0050 i& -0.10-0.0027 i\\
-0.57& 0.57& -0.02 i& 0.05 i& -0.02 i& -0.02 i& -0.22& 0.54& -0.00040\\                                                        
0.10 i& -0.07 i& -0.46& 0.46& -0.53& 0.53& 0.01 i& -0.01 i& -0.02 i\\                                                          
-0.00047& 0.000039& -0.01 i& -0.01 i& -0.01 i& -0.01 i& 0.71& 0.71& 0.000040\\                                                 
0.66& 0.33+0.0080 i& 0.30+0.0021 i& 0.30+0.01 i& 0.32+0.0019 i& 0.32-0.03 i& 0.18-0.0021 i& 0.18+0.0050 i& -0.10+0.0027 i\\
0.15 i& -0.13 i& -0.47& 0.57& 0.38& -0.47& 0.02 i& -0.02 i& 0.23 i\\                                                           
0.14-0.0084 i& 0.07-0.03 i& 0.41+0.05 i& 0.34-0.16 i& -0.38-0.06 i& -0.32+0.15 i& 0.04-0.00037 i& 0.03-0.02 i& 0.62\\          
\end{bmatrix}
\end{equation}

}

\ack{
This work was supported by the International Space Science Institute (ISSI) in Bern, through ISSI International Team project 457: The Role of Partial Ionization in the Formation, Dynamics and Stability of Solar Prominences.
We acknowledge valuable discussions with Prof. Andrew Hillier.
This work was supported by the FWO grant 1232122N and a FWO grant G0B4521N.
This project has received funding from the European Research Council (ERC) under
the European Union’s Horizon 2020 research and innovation programme (grant
agreement No. 833251 PROMINENT ERC-ADG 2018). This research is further supported by Internal funds KU Leuven, through the project C14/19/089 TRACESpace.
The resources and services used in this work were provided by the VSC (Flemish Supercomputer Center), funded by the Research Foundation - Flanders (FWO) and the Flemish Government.
}


\bibliographystyle{rusnat}

\end{document}